\documentclass[acmlarge]{acmart}
\renewcommand\footnotetextcopyrightpermission[1]{} 
\PassOptionsToPackage{hyphens}{url}\usepackage{hyperref}

\usepackage{booktabs} 

\usepackage{epsfig,endnotes}

\usepackage[english]{babel}
\usepackage{blindtext}

\usepackage{balance}
\usepackage[linesnumbered,ruled,vlined]{algorithm2e}

\SetAlFnt{\small}
\SetAlCapFnt{\small}
\SetAlCapNameFnt{\small}
\SetAlCapHSkip{0pt}
\IncMargin{-\parindent}

\usepackage{microtype}
\usepackage{caption}
\usepackage{subcaption}
\usepackage{indentfirst}
\newenvironment{edittext}{\color{black}}{}

\newcommand{\squishlist}{
 \begin{list}{${\bullet}$}
  { \setlength{\itemsep}{0pt}
     \setlength{\parsep}{3pt}
     \setlength{\topsep}{3pt}
     \setlength{\partopsep}{0pt}
     \setlength{\leftmargin}{1.5em}
     \setlength{\labelwidth}{1em}
     \setlength{\labelsep}{0.5em} } }
    
\newcommand{\squishend}{
  \end{list}  }

\acmConference{}{}{}
\begin{document}
\title{A System for Clock Synchronization in an Internet of Things}

\newcommand{\superscript}[1]{\ensuremath{^{\textrm{#1}}}}
\def\sharedaffiliation{\end{tabular}\newline\begin{tabular}{c}}
\def\wu{\superscript{*}}
\def\wp{\superscript{+}}
\def\wg{\superscript{\dag}}
\def\wk{\superscript{\#}}

\author{Sathiya Kumaran Mani}
\affiliation{%
 \institution{University of Wisconsin - Madison}
 \city{Madison}
 \state{Wisconsin}
 \country{USA}}
\email{sathiya@cs.wisc.edu}

\author{Ramakrishnan Durairajan}
\affiliation{%
 \institution{University of Oregon}
 \city{Eugene}
 \state{Oregon}
 \country{USA}}
\email{ram@cs.uoregon.edu}

\author{Paul Barford}
\affiliation{%
 \institution{University of Wisconsin - Madison}
 \city{Madison}
 \state{Wisconsin}
 \country{USA}}
\email{pb@cs.wisc.edu}

\author{Joel Sommers}
\affiliation{%
 \institution{Colgate University}
 \city{Hamilton}
 \state{New York}
 \country{USA}}
\email{jsommers@colgate.edu}



\maketitle
\thispagestyle{empty}
\renewcommand{\shortauthors}{S. Mani et al.}

\sloppypar

\subsection*{Abstract} \label{sec:abstract} 
Synchronizing clocks on Internet of Things (IoT) devices is important for
applications such as monitoring and real time control. In this paper, we describe 
a system for clock synchronization in IoT devices that is designed to be scalable, flexibly
accommodate diverse hardware, and maintain tight synchronization over a range
of operating conditions. We begin by examining clock drift on two standard IoT prototyping 
platforms. We observe clock drift on the order of seconds over relatively
short time periods, as well as poor clock rate stability, each of which make
standard synchronization protocols ineffective. To address this problem, we develop 
a synchronization system, which includes a lightweight client, a new packet exchange 
protocol called SPoT and a scalable reference server. We evaluate the efficacy of
our system over a range of configurations, operating conditions and target
platforms.  We find that SPoT performs synchronization 22x and 17x more accurately than MQTT and SNTP, respectively, at high noise levels, and maintains a clock accuracy of within $\sim$15ms at various noise
levels. Finally, we report on the scalability of our server implementation through microbenchmark and wide area experiments, which show that our system can scale to support large numbers of clients efficiently.

\section{Introduction}
\label{sec:introduction} Distributed Internet applications, including gaming, monitoring and real-time control, require that participating hosts have synchronized clocks.  Such synchronized clocks on network-connected devices enable shared experiences for users and coordination of application behaviors and interactions at specified times.  Clock synchronization between distributed devices is typically facilitated by interacting with a high-precision reference host that is deployed solely for this purpose.

%

Although the issue of clock synchronization in widely-distributed systems has been studied for many years ({\em e.g.}, see~\cite{Marzullo83,mills81dcnet,mills85alg}), Internet of Things (IoT) devices introduce several challenges.  First, an objective in IoT device design is to minimize costs.  This implies the use of lower quality hardware components including oscillators that generate clock signals on those devices.  Second, IoT devices are often deployed in environments with a broad range of operating temperatures. Low-quality clocks have been shown to run at widely varying rates depending on temperature~\cite{schmid2008exploiting}.  Finally, IoT devices often have limited computation and communication capability, which can constrain their ability to participate in standard clock synchronization protocols.

In this paper we consider the problem of synchronizing clocks in IoT devices
with a remote reference source.\footnote{This differs from the problem of
synchronizing clocks in a local deployment such as sensor networks, which may
not require synchronization with a global reference.} While NTP would be
a natural solution for clock synchronization, typical configurations require
stateful client computation and on-going communication with reference
source(s), which make Simple Network Time Protocol (SNTP) and similarly
lightweight mechanisms a more attractive choice.\footnote{We are not aware of
any NTP client implementations for IoT.} The goals of our work
are to understand how clocks operate on IoT devices and how they can be
synchronized in an accurate and efficient fashion.  The target platforms for our
experiments are the well known Arduino MKR1000~\cite{arduinomkr1000} and the
Raspberry Pi3~\cite{raspberrypi3}, both of which are often used for IoT
prototyping.  


We begin by examining the drift characteristics of our target IoT platforms
using raw millisecond counters. 
Our experiments on the different Arduino devices show inaccuracies
in synchronization ranging from 700 ms to as high as 1600
ms. Next, we test clock rate stability over a range of
temperatures that might be experienced for typical IoT device deployments.  We
observe clock drift as high as 600 ms over relatively short time
periods. Finally, we characterize the stability of the clock hardware
by measuring the {\em Allan variance}~\cite{mills1996networkAllan}. We find
that IoT clock hardware shows high variability and less stability
than traditional PC clock hardware. These results motivate the need for new synchronization
mechanisms that can accommodate lower clock stability and diverse clock drift
characteristics.

To improve synchronization of IoT devices,
we develop a new clock synchronization system that is designed to be scalable, lightweight and to enable synchronization on the order of 10 ms over a range of operating conditions.
The central components in our system are {\em (i)} a lightweight implementation
for IoT devices/clients, {\em (ii)} a scalable implementation of reference
servers that compute all key parameters ({\em e.g.}, clock offset, rate, etc.), and {\em (iii)} a packet exchange protocol that we call Synchronization Protocol for ioT (SPoT).  In our system, IoT clients simply contact a SPoT server and adjust their clocks to the value indicated in SPoT response packets with no additional computation required. \begin{edittext}This architecture allows SPoT to be used for time synchronization in many IoT deployment scenarios such as environmental sensing and early warning systems, smart homes, smart grids, smart vehicles, factory floor automation, gaming and IoT blockchain applications as described in  \S \ref{subsection:spotApplications}.\end{edittext}

The key technical challenge in our work is to develop clock synchronization
algorithms that are robust to the wide range of clock drift and offsets that
are expected in large IoT deployments.  We address this problem by developing
two novel algorithms:  one that synchronizes clock rates 
and one that addresses path asymmetry between client and server. 
The latter is developed in conjunction with a standard filtering method that allows packets with the best estimates to be selected.


We develop prototype implementations of the client and server components of our
system to evaluate its efficacy over a range of configurations, operating
conditions and target platforms. First, we
find that SPoT outperforms MQTT-based clock synchronization
mechanism and reports 16x, 19x, and 22x better clock offsets in the presence
of three different levels of noise.
Similarly, SPoT reported offsets are 3x, 10x, and 17x better than
SNTP. Next, we calculate the clock rate error ({\em i.e.}, Root Mean Squared
Error (RMSE) values) using NTP-reported offsets values and find that SPoT
consistently maintain accuracy within $\sim$15 ms at different noise levels.
Finally, we demonstrate the scalability of our server implementation in
a series of micro-benchmark and wide area tests.


In summary, this paper makes the following contributions.  First, we examine
the problem of synchronizing clocks on IoT devices by measuring clock drift on
two popular IoT prototype platforms over a range of operating conditions.
Second, we describe the design of a new, lightweight system for IoT clock
synchronization that is organized around a protocol called SPoT.  Finally, we develop
a reference implementation of our system\footnote{We will make all source code available upon publication.} and test it in a laboratory
and the wide area over a range of operating conditions.  We find that our system provides
much more accurate synchronization than standard protocols and that our server can scale to support a large number of clients.



\section{Background and Assumptions}
\label{sec:background} 
In this section we describe standard methods for network time synchronization, sources for errors, basics for IoT technologies, and our assumptions for devices that are the target for time synchronization. \begin{edittext}We also describe practical IoT deployment scenarios that could utilize our architecture to achieve accurate time synchronization.\end{edittext}

\subsection{Time Synchronization}

Synchronizing the clocks of networked devices is an important problem. Standard solutions include Network Time Protocol (NTP) (RFC 958~\cite{rfc958}) and its variants---including Simple Network Time Protocol (SNTP) (RFC 1769~\cite{rfc1769}), Precision Time Protocol (PTP)~\cite{ieeeptp}, Datacenter
Time Protocol (DTP)~\cite{lee2016globally} and Mobile NTP (MNTP)~\cite{Mani2016}, which have been developed for a variety of devices ({\em e.g.}, routers and wired hosts use NTP, mobile phones use SNTP/MNTP, datacenters use DTP, etc.). To effectively counter and correct the differing rate at which clocks advance (also known as {\em clock drift}), these protocols estimate one-way delays (OWDs) between the devices in which they are deployed and a hierarchy of clock references known as the {\em stratum servers}. The hierarchy starts with Stratum 0 servers, which are highly-precise clock sources
({\em e.g.}, GPS or atomic clock), and goes down to less-precise Stratum 15 servers.


Wired devices estimate {\em clock offsets} (defined as the difference in time between the client's clock and the remote reference) based on OWD measurements, which are determined using the timestamps exchanged with stratum servers in a process called {\em polling}. These timestamps are vetted using another process called {\em filtering}, which uses heuristics to remove inaccurate offset
estimates~\cite{ntpFilter, ntpHPFilter}. The filtering heuristics are optimized using indicators such as round-trip delay, jitter and oscillator frequency as part of the clock discipline algorithm~\cite{ntpClockDiscipline}. Both the clock filtering and synchronization algorithms are implemented and run as part of the {\tt ntpd} daemon in wired devices.

Similarly, wireless devices such as mobile phones use SNTP to acquire clock estimates by polling the stratum servers, but without NTP's sophisticated clock synchronization algorithm and filtering heuristics.  We also note that only the first octet of SNTP packets is set in the polling process employed by wireless devices, resulting in the exchange of less-accurate clock estimates.  Furthermore, wireless devices are known to employ vendor-specific SNTP implementations with varying polling frequencies and retry rates in case of polling errors/failures~\cite{Mani2016}.

\subsection{Clock synchronization errors}

Synchronizing a client's local clock with a reference time source consists of calculating two interdependent components: {\em (i) clock offset}; and {\em (ii) clock skew} or rate-error, which is the difference in rate or
frequency between the client's clock and the remote reference. While the offset synchronization ({\em i.e.}, calculating offset) is sufficient for general coordination of events among distributed clients, rate synchronization ({\em i.e.}, calculating clock skew) is necessary to achieve tight synchronization.
Surprisingly, while skew is a predominant source of synchronization error~\cite{moon1999estimation}, it is occasionally overlooked in distributed clock synchronization algorithms.  As we discuss in \S\ref{sec:sntp_evaluation}, inexpensive clock hardware is relatively unstable compared to traditional PC clock hardware and significant clock drift between synchronization requests exacerbates error. 

\subsection{Two-way synchronization method}

A fundamental operation in distributed clock synchronization is the
timestamped exchange of packets between a client and a reference, where exchanges are typically initiated by the client. This method, known as the \textit{two-way exchange}, often begins with the client sending a request packet at time {\em t1}. The server receives the packet at time {\em t2} and
sends a reply at time {\em t3}; the reception time of the reply at the client is {\em t4}. Timestamps {\em t1} and {\em t4} are from the client's clock, where as {\em t2} and {\em t3} are from the server's clock. The round trip time (RTT) of the exchange is given by $RTT = (t4 - t1) - (t3 - t2)$.

%

Some synchronization protocols ({\em e.g.}, SNTP) assume a symmetric forward \begin{edittext}({\em i.e.}, client-to-server)\end{edittext} and reverse \begin{edittext}({\em i.e.}, server-to-client)\end{edittext} delay. That is, they calculate OWD to be half of the RTT. Hence at true time {\em t2}, the client's time (T\textsubscript{client}) and offset with respect to {\em t2} are expressed as $T_{client}(t2) = t1 + owd$ and $offset(t2) = t2 - (t1 + owd)$.


The key takeaway is that the accuracy of the two-way exchange in calculating the clock offset at the client depends on validity of the assumption that the forward and reverse delays are symmetric and specifically that the forward delay is half of RTT. When this assumption is invalid due to path asymmetry, the delay gets added to the calculated offset as error. To illustrate the issue, consider an example in which the RTT of the exchange is 600 ms, with the forward delay being 400 ms and the reverse delay being 200 ms. If the client's clock is behind the reference by 20 ms, the correct offset that needs to be calculated from this exchange should be +20 ms. Because of the assumption of symmetric forward and reverse delays, however, the forward OWD is calculated as 300 ms with an underestimation error of 100 ms (which is half of the path asymmetry of 200 ms). This error in the calculation of the OWD is reflected in the offset calculation: +120 ms instead of the expected value of +20 ms. Similarly, if the forward and reverse delays are 200 ms and 400 ms, the calculated offset would be -80 ms due to the overestimation of the forward OWD by 100 ms. It can be shown that the error in offset calculation is bounded by 0.5 * RTT.

In wireline hosts, OWD asymmetry can occur due to path dynamism and variable switching delays, among other reasons. Such variability is often more pronounced in wireless hosts due to wireless
effects such as interference and channel noise~\cite{Mani2016}.  Given that the asymmetry error is bound by half of RTT, synchronization protocols typically address the error by preferring samples with smaller RTTs over others. Moreover,  synchronization protocols treat this as a {\em statistical variability problem}, hence their filtering approaches require multiple samples to pick the best RTT. For instance, NTP addresses this problem by collecting measurements from multiple reference servers and picking the server with the least RTT dispersion. From all the samples from that server, NTP selects the sample with the smallest RTT. Similarly RADClock~\cite{radClock04}, which has been shown to outperform NTP in terms of accuracy, maintains a moving window of measurements and uses a weighted sum of measurements within the window to calculate the offset.  Specifically, the samples with RTTs closer to the minimum RTT are weighted more than those that are much larger. Both protocols require multiple measurements for every offset calculation, extensive tuning, and in the case of the RADclock, a lot of state.  Approaches like Kalman filtering attempt to address these errors by modeling the clock in order to calculate the offset from noisy measurement samples. In addition to the difficulty of accurately modeling clock hardware, particularly for inexpensive hardware like those used by IoT devices, it is also computationally intensive, limiting scalability. In addition, Kalman filtering has been shown to exhibit degraded performance in the presence of non-Gaussian outliers~\cite{syncUnstableDelay16}.


\subsection{IoT Ecosystem}

{\bf IoT devices.} IoT devices are available for an increasingly wide variety of applications.  IoT devices differ in processing power, network connectivity, and packaging. Specifically, the processing capabilities range from microcontrollers running real-time operating systems to low-power (mobile) processors that are designed to function as gateways to connect to other IoT devices. In addition, IoT devices often use compact System-on-Chip (SoC) construction with a low-power radio chip to enable connectivity via cellular, WiFi, ZigBee, etc.  The SoC form factor enables transformation of standard devices ({\em e.g.}, thermostat) into ``smart" devices 
and the creation of entirely new types of devices such as wearables and voice assistants.


{\bf IoT cloud.} Current popular IoT cloud offerings, such as Amazon AWS IoT~\cite{awsiot} and Microsoft Azure IoT~\cite{azureiot}, are designed to serve billions of IoT device endpoints. 
IoT devices to communicate with services/applications running in the IoT cloud via protocols including  HTTP~\cite{rfc2616}, WebSockets~\cite{rfc6455}, AMQP~\cite{amqp} and MQTT~\cite{mqtt}. In particular, the widely-deployed MQTT protocol uses a publisher-subscriber model. IoT devices supporting MQTT can function both as publishers and subscribers of information. To coordinate the communication between publishers and subscribers, the MQTT protocol requires a {\em message broker}, which runs in the cloud and has the ability to scale with the number of cloud-connected IoT devices.

{\bf IoT time synchronization.}
IoT devices typically use SNTP or MQTT 
to synchronize 
with reference clocks. For the latter, either the cloud
platform pushes a timestamp over MQTT~\cite{pushEpoch} to the device similar to
the classic time protocol~\cite{rfc868}, or the device obtains a timestamp via
a standard REST API~\cite{pullRest} to synchronize with reference clocks.
In either case,
the IoT cloud server or message broker
publishes the current time to the IoT device {\em without accounting for
forward and reverse OWDs} of the packets carrying the timestamp and any response
message.
As such, \textit{any} OWD
contributes to synchronization inaccuracy at the device.  On the other hand, due to resource constraints in IoT devices, sophisticated OWD sample filtering heuristics
from the NTP protocol suite~\cite{ntpFilter} are too heavyweight and cannot be employed.  Furthermore, as a consequence of resource constraints in these devices, other variants of NTP ({\em e.g.}, PTP) are generally ill-suited for synchronizing devices' clocks in the IoT ecosystem.

\subsection{Assumptions}

Our system architecture assumes that: {\em (1)} IoT devices are capable of connecting to a cloud server, and {\em (2)} IoT protocols support heterogeneous protocol stacks. For example, some IoT devices are capable of connecting to the cloud directly over WiFi, while others use non-TCP/IP protocols ({\em e.g.,} ZigBee~\cite{zigbee} or Bluetooth Low Energy (BLE)~\cite{ble}) and thus require an IoT bridge to connect to the IoT cloud. Our architecture accommodates both WiFi-enabled IoT device that can connect directly to the cloud or that can function as a bridge for other IoT devices.

To support a wide range of IoT devices, we do not assume availability of unlimited compute power and/or processing capability.  Although our experiments target specific IoT platforms, we assume that the IoT device may have a very low-quality oscillator ({\em e.g.}, ceramic instead of crystal~\cite{arduinoclock}) and a limited energy source.  Lastly, we assume that IoT devices can respond to incoming packets and can adjust their clocks based on offset values sent to them. \begin{edittext}Our system currently assumes that devices are network-connected and can thus exchange synchronization messages with an IoT bridge or server.  In our ongoing work, we are examining how to efficiently and accurately synchronize devices that are intermittently connected, or may be disconnected over long duration.\end{edittext} Finally, we do not specifically address security issues in this paper.  Similar to other protocols, we assume that IoT devices will initiate time synchronization to a specified set of servers, which provides a minimal level of assurance. \begin{edittext}We posit that additional modes of packet exchanges, including secure hashes ({\em e.g.}, to detect man-in-the-middle attack) and encryption ({\em e.g.}, to enhance security) could be used without affecting the accuracy or correctness of our synchronization approach.\end{edittext} We intend to develop a more complete security model in future work.

\begin{edittext} 
\subsection{Applicability of SPoT for IoT deployments}
\label{subsection:spotApplications}
Given the design goals and the architecture of SPoT, we argue that it is applicable in a wide range of IoT applications. To provide a broad context and motivation for our work, we describe several deployment scenarios that would be able to utilize our architecture to achieve tight time synchronization.

\squishlist
 \item Consider a smart cloud manufacturing (S-CM)~\cite{Qu2016} shop floor environment that consists of arrays of sensors, controllers and automated assemblers/tools that build target widgets.  Key aspects of this environment are the ability to adapt the line on the fly for just-in-time assembly of custom, build-to-spec widgets. Such an environment requires tight clock synchronization between sensors, controllers and tools in order to assure quality and productivity and provide real-time monitoring. Further, to gain insights into telemetry data fused from globally distributed manufacturing sites, time synchronization with a global time reference such as coordinated universal time (UTC) becomes crucial. SPoT would be an ideal candidate for synchronizing the deployed IoT devices to UTC in such a scenario.
   \item Another example is earthquake detection IoT networks ~\cite{grillo, shakeAlert} that use low-powered accelerometers and cloud connected WiFi chips to detect and alert users about earthquakes in real-time. Tight time synchronization with a global time source would enable the development of sophisticated real-time earthquake detection algorithms, and given the quality of clock hardware found in such low-cost platforms and available cloud connectivity, the IoT network could utilize SPoT for accurate time synchronization.  
   \item  Yet another example is IoT deployments that utilize public blockchains to ensure trust and accountability among IoT devices that run distributed applications ~\cite{iotBlockchainSecurity, iotBlockchain, hyperLedger}.  Blockchain implementations require time synchronization among participating clients for protocol correctness~\cite{ethereumSyncError, paritySyncError} and preliminary code inspections reveal that popular blockchain implementations use SNTP~\cite{paritySNTP} which is shown to perform poorly under noisy network conditions (see \S \ref{subsec:spotvsothers} and ~\cite{Mani2016}). In such deployments, SPoT can facilitate accurate time synchronization among IoT nodes that participate in blockchain protocols.  
\squishend
\end{edittext}


\section{Time Synchronization in I\lowercase{o}T Ecosystem}
\label{sec:sntp_evaluation} 

In this section, 
we examine how the clocks on two popular IoT development platforms exhibit different drift characteristics based on the hardware instance and the ambient temperature. We characterize the stability of the clock hardware used by the IoT platforms and discuss the suitability of clock synchronization protocols from the wired/wireless Internet to the IoT ecosystem.

\subsection{Experimental setup}
\label{subsec:experimentalSetup}



{\bf IoT devices.} In our experiments we use four different Arduino MKR1000 devices (to compare instances of the same device), a commonly used platform for IoT prototyping~\cite{arduinoFamous}, and one Raspberry Pi3 (to contrast with the Arduino). The Arduino employs a SoC construction and uses a 32-bit low-power ARM MCU. It has a low-power 2.4 GHz 802.11 WiFi chip\footnote{To the best of our knowledge, this is the default connectivity option in IoT deployments.} to connect to the IoT cloud, and a crystal oscillator to drive the real-time clock which operates at 32.768 kHz. The Raspberry Pi3 uses a Broadcom BCM2837 SoC with a 1.2 GHz quad-core ARM Cortex A53 processor with 1GB LPDDR2-900 SDRAM. It supports 10/100 MBps Ethernet, 802.11n WiFi, and Bluetooth 4.0. In all our wide area experiments, we use Amazon's AWS IoT cloud, and the message broker is colocated with the cloud server.



{\bf Testbed.} Our experimental testbed
is designed to examine drift characteristics of the IoT platforms. The testbed includes three components: {\em (1)} a wireless access point, {\em (2)} IoT device, and {\em (3)} a monitor node (Macbook pro laptop). The IoT device connects to the IoT cloud through the wireless access point. The monitor node is connected to the IoT device over a serial port interface and is used to collect timestamps (and other relevant statistics) from the device. In the experiments described below, the NTP-corrected system clock of the monitor node is used as the reference clock to benchmark the internal system clock of the IoT device. That is, the monitor node is synchronized with {\tt 0.pool.ntp.org} before the start and throughout the duration of experiment, which is 1 h. We run the experiment for 24 h to collect clock offset (for all devices), which we use in our trace-driven analysis for testing SPoT's scalability (\S\ref{sec:iotntp_evaluation}).



\subsection{Drift characteristics of clock hardware}

\label{temperature_experiments}

To gain perspective on IoT clock synchronization we examine the drift
characteristics of different Arduino hardware instances. 
While we only consider a single type of device in these experiments, it is a popular and widely used platform which we believe provides a realistic representation of IoT device capabilities and characteristics. We also consider clock drift under different ambient temperature conditions.

Our experiments gather a pair of timestamps every second. The first timestamp is obtained using the NTP-\begin{edittext}disciplined\end{edittext} system clock of the monitor node. The second is a {\em raw} millisecond counter value obtained from the IoT hardware's clock over the serial port. We then calculate clock offsets using the timestamps measured from the IoT device's raw counter and from the NTP-corrected monitor node. We run the experiments 
at three different locations:
{\em (a)} {\em L1}: a temperature-controlled server room maintained at $\sim$14
$^{\circ}$C, {\em (b)} {\em L2}: an office room setting at $\sim$21
$^{\circ}$C, and {\em (c)} {\em L3}: a residential apartment at $\sim$27.5
$^{\circ}$C.  We consider these locations and their respective temperatures as
representative of common real-world IoT deployments.

\begin{figure*}[htb!]
 \includegraphics[width=7.8cm]{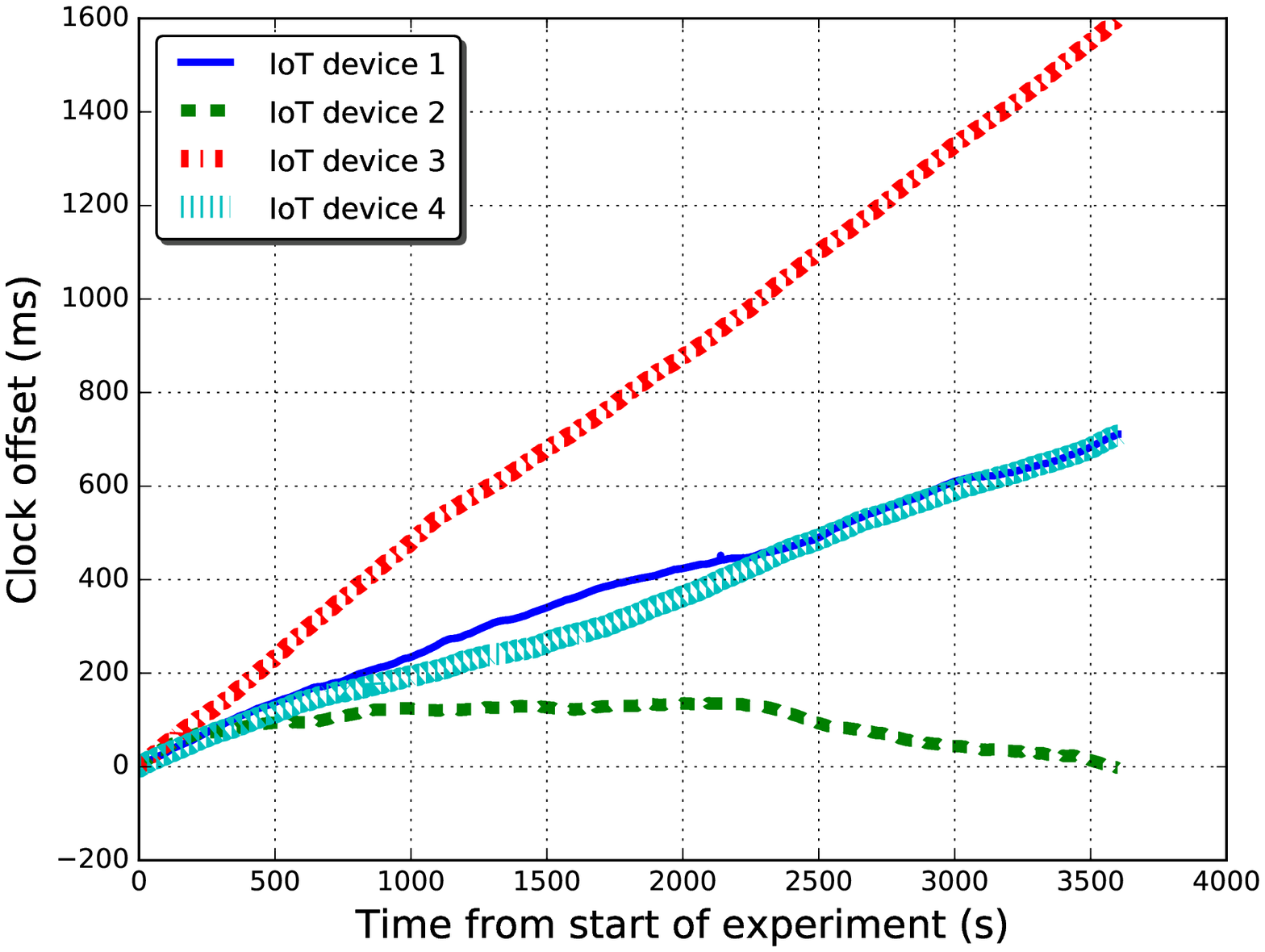}
 \includegraphics[width=7.8cm]{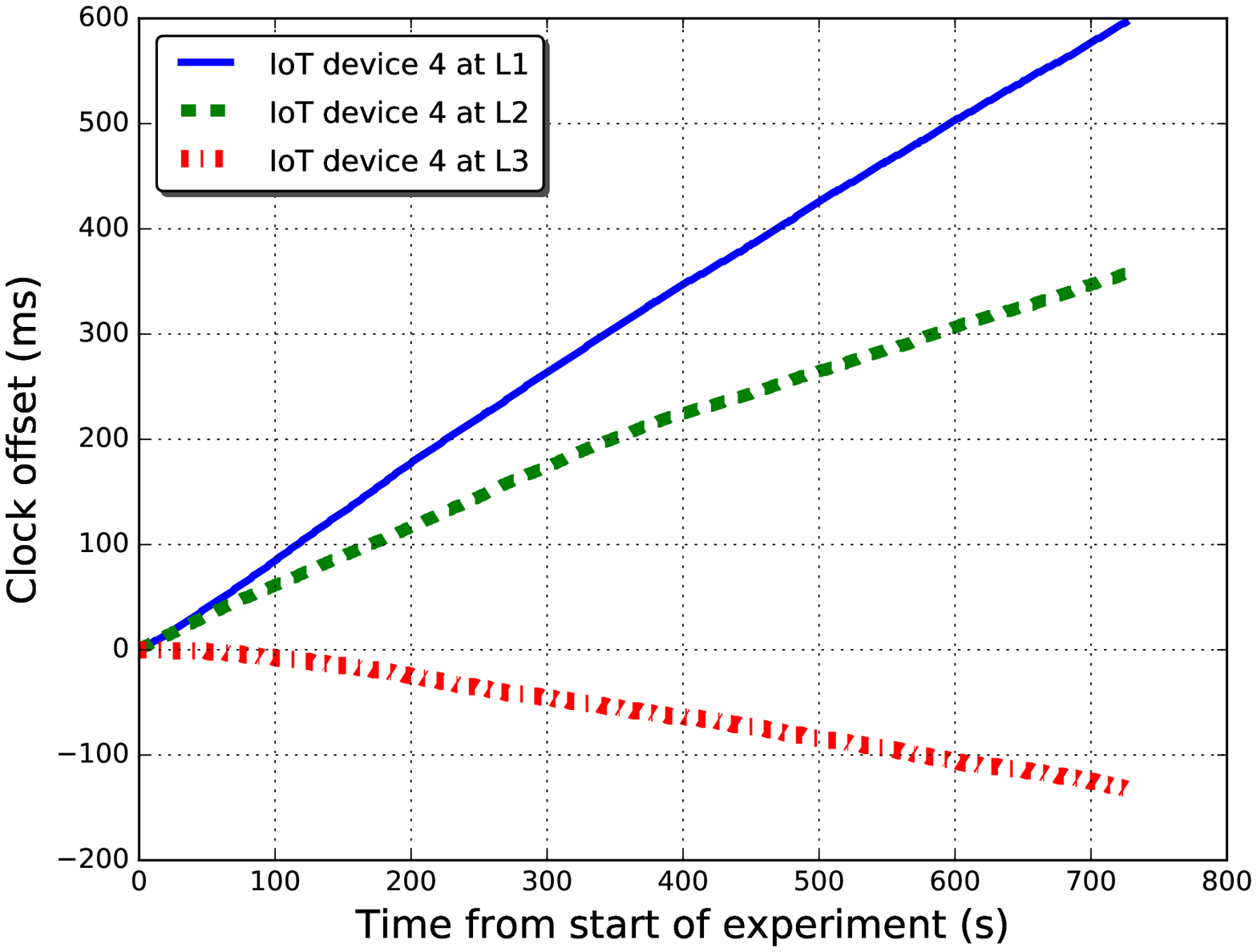}

 \caption{\label{fig:singleAndMultiDeviceOffsets}{\bf Comparison of drift
characteristics of different Arduino devices (left) and same Arduino device under
different ambient temperatures (right).}}
\end{figure*}

Figure~\ref{fig:singleAndMultiDeviceOffsets} shows the calculated clock offsets throughout the experiment for different hardware instances at locations L2 (left) and clock offsets of the same hardware instance at all three locations L1, L2 and L3 (middle). Visually, we observe that different hardware instances of the same prototyping platform show quite different drift characteristics and drift rates and hence different values of clock offset for the same duration of the experiment. Further, even the same hardware instance exhibits different clock drift rates depending on the ambient temperature.  The plots show that the difference in behavior due to the differences in ambient temperature is apparent even for small durations ({\em i.e.}, 10 min).

{\bf Discussion.} While the differences in manufacturing, environmental conditions and crystal aging leading to variations on drift characteristics of the same type of hardware are well known, here we observe that the variations are large for inexpensive clock hardware used by a canonical IoT device. Also, the amount of clock drift is huge (about $\sim$600 ms for a period of 10 min) in
the worst case, compared to traditional PC clock hardware~\cite{ridoux2012case,veitch2009robust}. This variability in drift characteristics
implies that {\em accurate synchronization will be challenging for any mechanism that expects a uniform drift characteristic from all IoT devices}. Moreover, for mechanisms that expect to model the clock hardware and that involve the use of training data, we posit that clock synchronization will be resource intensive due to the need to model/train on a wide range of drift characteristics at different ambient temperatures~\cite{kim2012tracking,hamilton2008aces}.

\subsection{Stability of clock hardware}
\label{stability_experiments}

Next, we compare the stability of clock hardware on the Arduino devices with the Raspberry Pi3. Our goal is to understand if the assumptions made in the design of several wired/wireless time synchronization protocols are broadly applicable.  A typical measure of oscillator stability (and hence clock stability) is the {\em Allan variance}~\cite{mills1996networkAllan}. Given a series of consecutive offset measurements of a clock at certain measurement interval $\tau$, the fractional frequency or instantaneous clock skew is the rate of change of offset calculated using consecutive offset
measurements. The Allan variance is an estimator of the variance of the clock skew for the given measurement interval.
Since the rate stability of a clock is a function of $\tau$, often the square root of the Allan variance called the {\em Allan deviation} is plotted as a function of $\tau$ on a log-log curve to study the stability of any given clock~\cite{radClock04}.

The Allan deviation curve of typical PC clock hardware when viewed on a log-log plot is often made up of two lines: {\em (a)} the line with a slope of -1, dominated by white phase noise, which is the measurement noise characteristic of the channel; and {\em (b)} the line with slope of +0.5, dominated by the random walk noise, which is characteristic of the clock wander. The intersection of these two
lines---which is also the inflection point in the curve---is called the {\em Allan intercept}. The Allan intercept characterizes the given clock hardware and network path: that is, it represents the measurement interval where the error due to the measurement noise from the network path and the frequency error due to the wander of the clock would be minimal. Thus, it is an important
statistic that influences the design of rate synchronization and polling behavior of many time synchronization protocols such as NTP~\cite{ntpClockDiscipline98} and RADClock~\cite{radClock04}. These protocols expect to see the intercept in the range of $\tau$ = 1000 s.

\begin{figure*}[htb!]
\includegraphics[width=8cm]{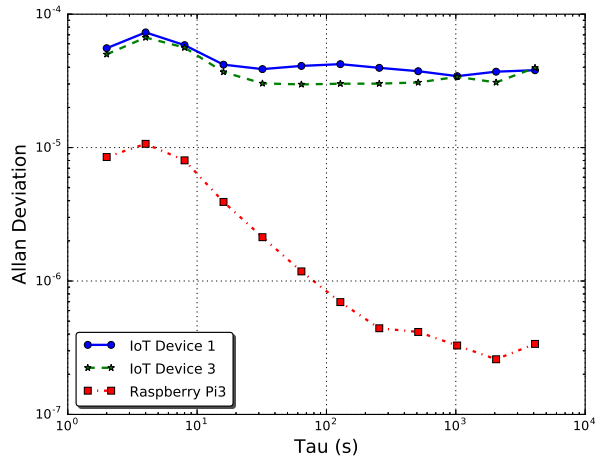}

 \caption{\label{fig:allanDeviationplot}{\bf Comparison of Allan deviation plots of Arduino devices 1 and 3 and Pi3.}}
\end{figure*}

Figure~\ref{fig:allanDeviationplot} shows the Allan deviation plots for the Raspberry
Pi3 and two Arduino devices ({\em i.e.}, IoT devices 1 and 3) used in our experiments, all using offset
measurements collected at location L2. The shape of the Allan deviation curve and the Allan intercept around  ~1000 s for Pi3 is consistent with prior studies\cite{timeToMeasurePi}. From the figure, it is
clear that the stability of Arduino clock hardware is much lower and does not satisfy the stability assumptions made by other clock synchronization protocols. The Allan deviation is in between $10^{-4}$ and $10^{-5}$, that is, the variability of the clock frequency/rate is in the range of 10s of ms even for short measurement intervals, which is consistent with our earlier observations (see Figure~\ref{fig:singleAndMultiDeviceOffsets}).  Also the flatter Allan deviation curves indicate that the clock wander effects start to dominate much earlier that 1000 s. It should be noted that the Allan deviation of Pi3 is slightly higher than the one seen in~\cite{timeToMeasurePi}, which is simply because of our measurement setup where we measure Pi3 offsets from across a LAN using the monitor node in our setup discussed in \S\ref{subsec:experimentalSetup}---an observation consistent with prior
studies~\cite{ntpClockDiscipline98}.

%

{\bf Discussion.} From Figure~\ref{fig:allanDeviationplot} it is clear that IoT clock hardware shows high variability and are less stable than traditional PC clock hardware. In addition, it is evident that IoT hardware do not meet the design assumptions used by synchronization protocols like NTP, RADClock, etc., signaling the need for synchronization mechanisms that can accommodate lower
clock stability and diverse clock drift characteristics. This lower observed stability
also implies that the estimation of Allan intercept as part of the synchronization mechanisms would be quite challenging and calls for simpler methods to estimate the stability of the clock and to pick suitable measurement/polling intervals. These insights are key to our proposed approach for rate synchronization and varying the polling interval.


\section{SP\lowercase{o}T Design}
\label{sec:iotntp_design} In this section, we describe the design objectives, organization and implementation of our system for clock synchronization. We also describe the details of the two key algorithms that enable tight clock synchronization on IoT devices.


\subsection{SPoT Architecture}

\label{sec:arc}

Our system for clock synchronization is designed to be applicable to IoT devices that vary in terms of computation capabilities, clock hardware stability, energy budget and accuracy requirements. Our synchronization algorithms are designed to be lightweight in terms of packets exchanged and computations involved as well as state information maintained.   Our system is also designed to scale to support large numbers of devices that synchronize with reference servers so that clocks are consistent within 10s of milliseconds.  The system is also designed to be flexible and extensible so that individual components such as filtering algorithms or packet exchange protocols can be updated as improvements become available.

The basic design components include software that runs on clients and a server infrastructure that runs in a cloud environment.  To support IoT devices with different computational capabilities, our architecture allows two types of synchronization clients namely, thick clients and thin clients. Thick clients run the lightweight synchronization algorithms with the polling interval between synchronization requests, which are determined by the algorithms. The server simply responds to the timestamped message exchanges initiated by these clients. For thin clients with limited compute capabilities, the
synchronization algorithms are run on the server. Thin clients are expected to respond to the timestamped message exchange requests sent from the server and use the offset and the clock skew provided by the server.  Further, the energy limitations of the IoT devices can be addressed by choosing
different polling regimes and by setting the Error Margin (EM) accordingly. In short, our architecture operates as follows:

\squishlist
  \item IoT devices register with SPoT server with information such as device
  type, polling style and EM.
   \item  Thin clients respond to the timestamp exchanges initiated by the SPoT
     server. They use the offset and skew values provided.
   \item  Thick clients execute the synchronization algorithms and exchange
     timestamped messages with SPoT server as determined by the algorithms.
\squishend

Beyond the core components, our system includes algorithms for offset synchronization and rate synchronization, which we describe below.

\subsection{Offset synchronization in SPoT}

\label{subsec:offset}

The key to achieving good offset synchronization is to directly address asymmetry errors.
This is accomplished by identifying the \textit{direction} ({\em i.e.}, is the asymmetry on the forward or the reverse path?) and \textit{magnitude} of the asymmetry.
Once these have been identified,
SPoT corrects for the asymmetry error in the offset calculation. To identify the direction of the asymmetry for a given measurement, we consider how the offset is affected by a particular measurement. Given an expected offset, from the earlier discussion, it is clear that an asymmetry in the forward direction \begin{edittext}({\em i.e.}, client-to-server)\end{edittext} increases the expected offset by an amount equal to half of the magnitude of asymmetry; similarly, an asymmetry in the reverse direction \begin{edittext}({\em i.e.}, server-to-client)\end{edittext} decreases the offset by an amount equal to half the magnitude of the asymmetry.  Hence by comparing the computed offset to the expected offset, the direction of the asymmetry can be inferred. This insight forms the basis of SPoT's filtering
approach. The magnitude of the asymmetry is estimated to be the difference between the minimum RTT of all samples seen so far and the RTT of the current sample.

SPoT uses Algorithm~\ref{alg:SPoT_filter} to correct the error introduced by asymmetric path delays. Given the clock skew and last known offset and its measurement time, the algorithm calculates an estimate of the current clock offset (step 2) and the magnitude of the asymmetry (step 3). If the
measured offset is significantly greater than the estimated offset, the asymmetry is determined to be on the forward path and the measured offset is corrected accordingly (step 5). Similarly, correction is also applied for reverse path asymmetry (step 7). If both the tests are not satisfied, the additional
delay is inferred to be a symmetric additional delay and the measured offset is
accepted without any correction (step 9). The test to check if the measured
offset is significantly greater or lesser than the estimated offset is based on
a user tunable threshold called the EM, which is set to 10ms in our experiments.\footnote{This threshold was determined experimentally and proved to be robust across our evaluations.  We do not include a sensitivity analysis due to space limits.}

\SetKwProg{Fn}{Function}{}{}
\begin{algorithm}[htb!]
\SetKwInOut{Input}{input}
\newcommand{\cmtsty}[1]{\texttt{\small #1}}\SetCommentSty{cmtsty}
\Input{$measuredOffset$ = offset from measurement}
\Input{$measuredRTT$ = RTT of measurement}
\Input{$oldOffset$ = last known offset}
\Input{$timeDelta$ = duration since last measurement}
\Input{$clockSkew$ = estimated rate-error of clock}
\Input{$minRTT$ = minimum RTT seen}
\Input{$errorMargin$ = tunable error margin}
\Fn{filterOffset()}{
	$estimatedOffset$ = oldOffset + clockSkew * timeDelta\\
	$asymmetricDelay$ = measuredRTT - minRTT\\
	\If{$measuredOffset$ \textgreater  $estimatedOffset$  + $errorMargin$}{
		\tcp{forward asymmetric error}
		$correctedOffset$ = $measuredOffset$ - 0.5 * $asymmetricDelay$\\
	}
	\ElseIf{$measuredOffset$ \textless  $estimatedOffset$  - $errorMargin$}{
		\tcp{reverse asymmetric error}
		$correctedOffset$ = $measuredOffset$ + 0.5 * $asymmetricDelay$\\
	}
	\Else{
		\tcp{symmetric additional delay}
		$correctedOffset$ = $measuredOffset$\\
	}

	\Return $correctedOffset$\\
}

\caption{Offset synchronization algorithm}
\label{alg:SPoT_filter}
\end{algorithm}

\subsection{Rate synchronization in SPoT}

\label{subsec:rate}

An important input in our system is the clock skew, which is required for calculating an estimated clock offset.  For devices with lower-quality clock hardware that can drift significantly between synchronization points (see \S\ref{sec:sntp_evaluation}), the clock skew is also necessary for correcting for the clock drift when reading time on the device between synchronization points.  Estimating and updating the clock skew is the problem of {\em rate synchronization}. SPoT's rate synchronization algorithm works in conjunction with its offset estimation algorithm and is shown here separately for clarity. 

The accuracy of the calculated clock skew depends on the stability of the clock hardware and the duration between subsequent measurements. For stable clock hardware, the calculated clock skew will remain valid and accurate for longer durations and hence the polling intervals could be large. For hardware that is less stable, the clock skew should be updated more often and hence the reference should be polled more frequently. Further, the accuracy of the clock skew has an impact on the accuracy of the offset algorithm.  SPoT's rate synchronization algorithm, which is shown in Algorithm~\ref{alg:SPoT_rate_sync}, uses this insight to determine the stability of the clock hardware and picks the best possible polling interval between measurements.

\SetKwProg{Fn}{Function}{}{}
\begin{algorithm}[htb!]
\SetKwInOut{Input}{input}
\newcommand{\cmtsty}[1]{\texttt{\small #1}}\SetCommentSty{cmtsty}
\Input{$measuredOffset$ = offset from measurement}
\Input{$errorMargin$ = tunable error margin}
\Input{$pollingStyle$ = AIMD or MIMD as chosen}
\Fn{synchronizeClockRate()}{

	\If{In observation time or $numSamples$ \textless 5}{
        $absoluteError$ = abs($estimatedOffset$ - $correctedOffset$)\\
        $meanAbsoluteError$.update(absoluteError)\\
		$numSamples$ += 1\\
	}
	\ElseIf{$meanAbsoluteError$ \textless  2 * $errorMargin$}{
		\tcp{clock has been stable so far}
		\tcp{increase polling interval}
		$pollingInterval$.increase($pollingStyle$)\\
		restartObservationTime()\\
		$numSamples$ = 0\\
	}
	\Else{
		\tcp{clock is unstable}
		\tcp{decrease polling interval}
		$pollingInterval$.decrease($pollingStyle$)\\
		restartObservationTime()\\
		$numSamples$ = 0\\
	}
	\If{$correctedOffset$ = $measuredOffset$}{
		\tcp{high quality offset sample}
		\tcp{update clock skew}
		$clockSkew$ = ($clockSyncOffset$ - $correctedOffset$) / ($clockSyncTime$ - currentTime())\\
		$clockSyncOffset$ = $correctedOffset$\\
		$clockSyncTime$ = currentTime()\\
	}
}

\caption{Rate synchronization algorithm}
\label{alg:SPoT_rate_sync}
\end{algorithm}

The rate synchronization (Algorithm~\ref{alg:SPoT_rate_sync}) is run every time the offset algorithm is executed. The frequency of running the offset algorithm and hence the frequency of polling or measuring the clock offset is controlled by the polling interval as determined by the rate synchronization algorithm. Algorithm~\ref{alg:SPoT_rate_sync} calculates the absolute error between the estimated offset and the corrected offset for every offset synchronization point. For an observation time (set to 5 minutes in our experiments) or at least until 5 such absolute errors have been observed, which
ever is longer, the algorithm calculates a running mean of these absolute errors (step 4). Once the observation time has expired, the mean absolute error is compared against the EM. If the mean absolute error is less than twice the EM, the clock is determined to be stable and hence the polling interval is increased (step 7). Similarly, if the mean absolute error is greater than twice the EM, the clock is deemed to be unstable and the polling interval is decreased accordingly (step 11). Further, higher quality samples, where the measured offset and corrected offset are same, are used to update the clock skew. The amount of increase or decrease applied to the polling interval depends on the polling style chosen by the user. 

The polling style could be Additive Increase and Multiplicative Decrease (AIMD) or Multiplicative Increase and Multiplicative Decrease (MIMD) depending on the accuracy requirements and energy budget of the device. For devices with lower energy budget ({\em e.g.}, operating on battery power) and lower accuracy requirements, MIMD could be used. MIMD is aggressive in increasing the polling interval in order to reduce the number of offset synchronization measurements for slightly reduced
accuracy. EM could also be increased by the user to further decrease the polling cost incurred by the synchronization algorithms in lieu of synchronization accuracy.

\subsection{SPoT Implementation}

As described in \S\ref{sec:arc}, SPoT is designed to support two types of clients: thick and thin. Thick clients run the entire offset and rate synchronization algorithms, whereas for thin clients the synchronization algorithms are run {\em only on the server}. To accommodate both clients, SPoT
implementation consists of four major components: {\em (1)} the core library implementing synchronization algorithms for thick clients, {\em (2)} the scalable reference server implementation that supports both thick clients and thin clients,  {\em (3)} a reference implementation for a thin client, and {\em (4)} a client emulator that can support multiple thin clients on a single physical node, used in our scalability experiment (see \S\ref{sec:iotntp_evaluation}). The core library is implemented in $\sim$400 lines of C. It can be directly used by thick clients and by the reference server implementation to support thin clients.

The SPoT synchronization library is designed to be lightweight even for thick clients and maintains only 15 variables to manage the synchronization, two of which are the offset and rate estimates.  SPoT's lightweight implementation makes it well-suited for resource-constrained IoT platforms, some of which do not allow tasks to run in a dedicated thread~\cite{particleThreading}.  Existing synchronization systems that run in dedicated daemon processes~\cite{ntpd, radClock04} would simply not be possible on such platforms.

The reference server is implemented in $\sim$300 lines of C++ using an asynchronous networking library~\cite{asioLib} and is capable of supporting thick and thin clients.  SPoT server support
for thick clients is stateless, since it only needs to provide timestamps t2 and t3 in any packet exchange.
The server uses multiple threads: one to send timestamp exchange with thick clients and
device registration, one scheduler thread to initiate timestamp exchange with each of the registered thin clients based on the polling interval selected by the synchronization algorithm for each client and several threads to manage the asynchronous sending and reception of the timestamp probes with thin clients.


The client emulator is also written in C++ in $\sim$130 LoC. The client emulator runs multiple thin clients and responds to the timestamp probes from the server.

\section{SP\lowercase{o}T Evaluation} 
\label{sec:iotntp_evaluation} In this section, we evaluate our system by comparing the clock synchronization
accuracy of SPoT with other protocols. We also examine the polling behavior of SPoT and perform microbenchmarks to assess the scalability of our system.

\subsection{Experimental Approach}
\label{subsec:compareSync}

To evaluate SPoT, we use the experimental testbed described in
\S\ref{subsec:experimentalSetup} as follows. First, we collect clock offsets from the two Arduino-based IoT devices for a period of 24 hr. Next, to test the synchronization accuracy of SPoT and to compare SPoT against other protocols such as SNTP, MQTT, and filtering methods used by MNTP~\cite{Mani2016} and {\tt ntpdate}~\cite{ntpdate}, we add {\em observational noise} to the collected offsets and perform trace driven analysis. \begin{edittext}We compare SPoT with these protocols due to their widespread adoption in the Internet~\cite{Mani2016} and the preference among developers to use these protocols for a surprisingly wide range of scenarios from popular mobile applications~\cite{iosSNTP, snapchatSNTP} to blockchain clients~\cite{paritySNTP}.\end{edittext} We use a trace driven approach since the addition of noise in live tests is extremely difficult to control. The observational noise is drawn from a normal distribution with zero mean and a given standard deviation, which we set at different levels in our experiments. 
To ensure that both noisy and noiseless offset observations are equally probable, we add the observational noise to each offset measurement in the trace with probability 0.5. To generate RTT measurements corresponding to the noisy offset measurements, we begin with the observation that the two-way method halves the actual noise and the actual noise is simply additive delay on top of the path
RTT. That is, we start with a given path RTT and add twice the absolute value of the corresponding observational noise as an additive delay. The path RTT which is set to 300 ms\footnote{We selected this value to approximate the RTT between an IoT device and a cloud-based reference server.} in our analysis can be set to any nominal value without affecting the results of the analysis. \begin{edittext}We do not consider network errors such as packet drops, duplicates etc., since the mechanisms to handle these errors such as retries and timeouts are implementation details that do not affect the correctness or accuracy of our algorithms. We do not consider scenarios like extended periods of loss of connectivity and plan to address this as part of our future work.\end{edittext}

Using the traces, we compare SPoT against other protocols in three levels of noise: low, medium, and high; where the standard deviations of the noise distribution are set to 50 ms, 150 ms, and 250 ms respectively.
We run the different synchronization protocols on the 24-hr trace and collect statistics including RMSE, minimum, maximum, and standard deviation of offset errors. In our results, we report error statistics averaged over 100 runs. In all of our experiments, we run SPoT with AIMD polling behavior and with error margin set to 10 ms. Moreover, in addition to comparing SPoT with widely-used IoT-specific synchronization protocols, we also compare SPoT with two other mechanisms.
(1) Consensus is the offset filtering method used in prior work~\cite{Mani2016} for bootstrapping the synchronization mechanism with high quality measurements. To report each offset, the Consensus method makes 8 measurements, each 15 s apart. Outliers are eliminated and the average of the remaining measurements are used. (2) MinRTT is the filtering approach used by the standard {\tt ntpdate} implementation where an offset measurement with a minimum RTT among 8 samples is selected~\cite{ntpdate}. \begin{edittext}In our evaluations we do not compare thick and thin implementations since they offer the same accuracy. The only difference between the two implementations is computational cost incurred by thick clients, which is offloaded to the SPoT server in case of thin clients.\end{edittext}

\subsection{SPoT vs. other protocols}
\label{subsec:spotvsothers}

{\bf Offset errors.} Table~\ref{tab:rmseValues} compares the RMSE among different synchronization
protocols at different noise levels. The values in parenthesis are the
RMSE values obtained from raw offset measurements without any filtering. Since
SNTP and MQTT have no filtering capabilities, their offsets are
the same as raw offsets. From the table we observe that SPoT performs
consistently well by maintaining the same level of accuracy ($\sim$10 ms), even under
high noise levels, where as the accuracy of other protocols deteriorate with
increasing noise levels. Specifically, SPoT performs 16x, 19x, and 22x better than
MQTT in low, medium, and high noise levels. Similarly, SPoT provides 3x (low), 10x
(medium) and 17x (high) better accuracy versus SNTP. Of all the protocols examined, MQTT
has the lowest clock synchronization accuracy followed by SNTP.  Although
Consensus and MinRTT methods perform well with low noise levels,
their accuracy is affected by increases in noise levels.

\begin{table}[htb!]
\centering
\caption{{\bf Comparison of synchronization RMSE (ms) values at different noise levels.}}
\label{tab:rmseValues}

\begin{tabular}{c|c|c|c}
Protocol & Low noise & Medium noise & High noise \\ \hline
SPoT & 10.0 (35.6) & 9.3 (102.1) & 8.9 (173.4) \\
SNTP & 36.2 (36.2) & 105.5 (105.5) &  161.7 (161.7)\\
MQTT & 162.5 (162.5) & 196.6 (196.6) & 225.7 (225.7) \\
Consensus & 9.1 (34.5) & 27.0 (109.8) & 42.5 (175.7) \\
MinRTT & 8.7 (34.8) & 21.8 (107.0) & 41.9 (177.5)
\end{tabular}

\end{table}

Figure~\ref{fig:1hrPlot} shows the offsets reported by SPoT for the first hour of the experiment under high noise level for IoT devices 1 (top) and 2 (bottom). From this figure, we first observe that SPoT is effective in correcting offset measurements in the face of high noise levels, whereas the unfiltered offsets produced by SNTP are as high as 600 ms. Second, we see that clock skew estimates produced by SPoT's rate synchronization algorithm follow the original ground truth offset, despite the different non-linear clock drift trends of devices.

We also compared the performance of SPoT, MNTP, and SNTP using a separate trace-based experiment.  In this experiment, we assume the client is running in an IoT bridge environment since MNTP makes certain assumptions about clock drift that make it inappropriate to use directly on an IoT device that may exhibit non-linear drift~\cite{Mani2016}.  The trace used was of SNTP packet exchanges every 5s, used in our prior work~\cite{Mani2016}.  Since SPoT uses a variable polling interval, we interpolated the raw SNTP measurements by using the same value for any time during a given 5s interval.  The maximum synchronization RMSE (ms) for SPoT, MNTP, and SNTP, was 0.72, 6.172, and 51.89, respectively, showing that SPoT's algorithms provide a significant boost in accuracy over the other techniques.

\begin{figure}[htb!]
  \centering
  \includegraphics[width=14cm]{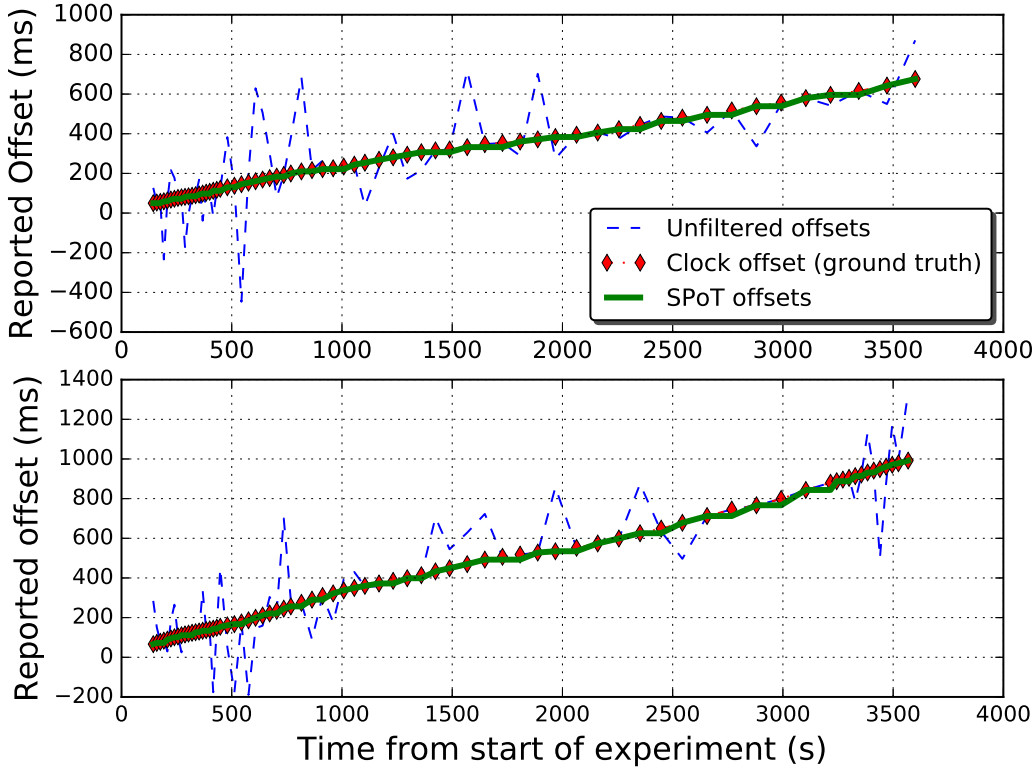}
  \caption{\label{fig:1hrPlot}{\bf Offsets of IoT devices 1 (top) and
2 (bottom) reported by SPoT from the first hour
of the experiment with high noise level.}}
\end{figure}

{\bf Error statistics.} To complement Figure~\ref{fig:1hrPlot}, Table~\ref{tab:errorStats} shows the minimum, maximum and standard deviation of offset errors for the different synchronization protocols for high noise level. We make two key observations: {\em (1)} since MQTT is a push-based mechanism that simply publishes the timestamps to IoT clients, a major source of error is the uncorrected OWD in the published timestamps; and {\em (2)} none of the synchronization mechanisms, except SPoT, provide rate synchronization and simultaneous variation of polling interval based on stability of device's clock ({\em i.e.}, they run with a default polling value of 128 s). The minimum error of 0 ms for all protocols except MQTT is due to their ability to identify/use noiseless measurements in/from a noisy environment. While Consensus and MinRTT protocols reduce the maximum error with respect to the
raw unfiltered offset measurements, it is clear that SPoT is more effective: it bounds the maximum error to be within 50 ms. Furthermore, the standard deviation of offset errors for SPoT is lower than other protocols ({\em i.e.}, within 10 ms).

\begin{table}[htb!]
\centering
\caption{{\bf Comparison of offset error statistics under high noise level.}}
\label{tab:errorStats}
\begin{tabular}{c|c|c|c}
Protocol & Minimum (ms) & Maximum (ms)  & Standard Deviation (ms) \\ \hline
SPoT & 0.0 (0.0) & 47.5 (771.36) & 7.8 (142.4) \\
SNTP  & 0.0 (0.0) & 709.0 (709.0) & 133.8 (133.8) \\
MQTT & 150.0 (150.0) & 781.7 (781.7) & 107.6 (107.6) \\
Consensus & 0.0 (0.0) & 253.1 (896.6) & 32.3 (144.9) \\
MinRTT & 0.0 (0.0) & 417.7 (855.1) & 40.2 (147.6)
\end{tabular}
\end{table}

Next, we calculate the rate errors at each synchronization point by estimating an offset using the clock skew provided by SPoT in comparison with the ground truth offset value. Hence these errors provide a bound for worst case offset errors incurred by using the clock skew estimates that are produced by the rate synchronization process.  We observe that SPoT's rate synchronization is able to achieve RMSE values of 14.7 ms, 13.3 ms and 13.5 ms under conditions of low, medium and high noise. It is clear that SPoT's rate synchronization accuracy is consistent under all noise levels ({\em i.e.}, rate errors are less than 15 ms). Since the rate synchronization mechanisms vary the polling interval depending on the stability of the clock hardware, we note that SPoT's polling is adaptive and robust against all noise levels.

\begin{figure*}[htb!]
  \centering
  \includegraphics[width=14cm]{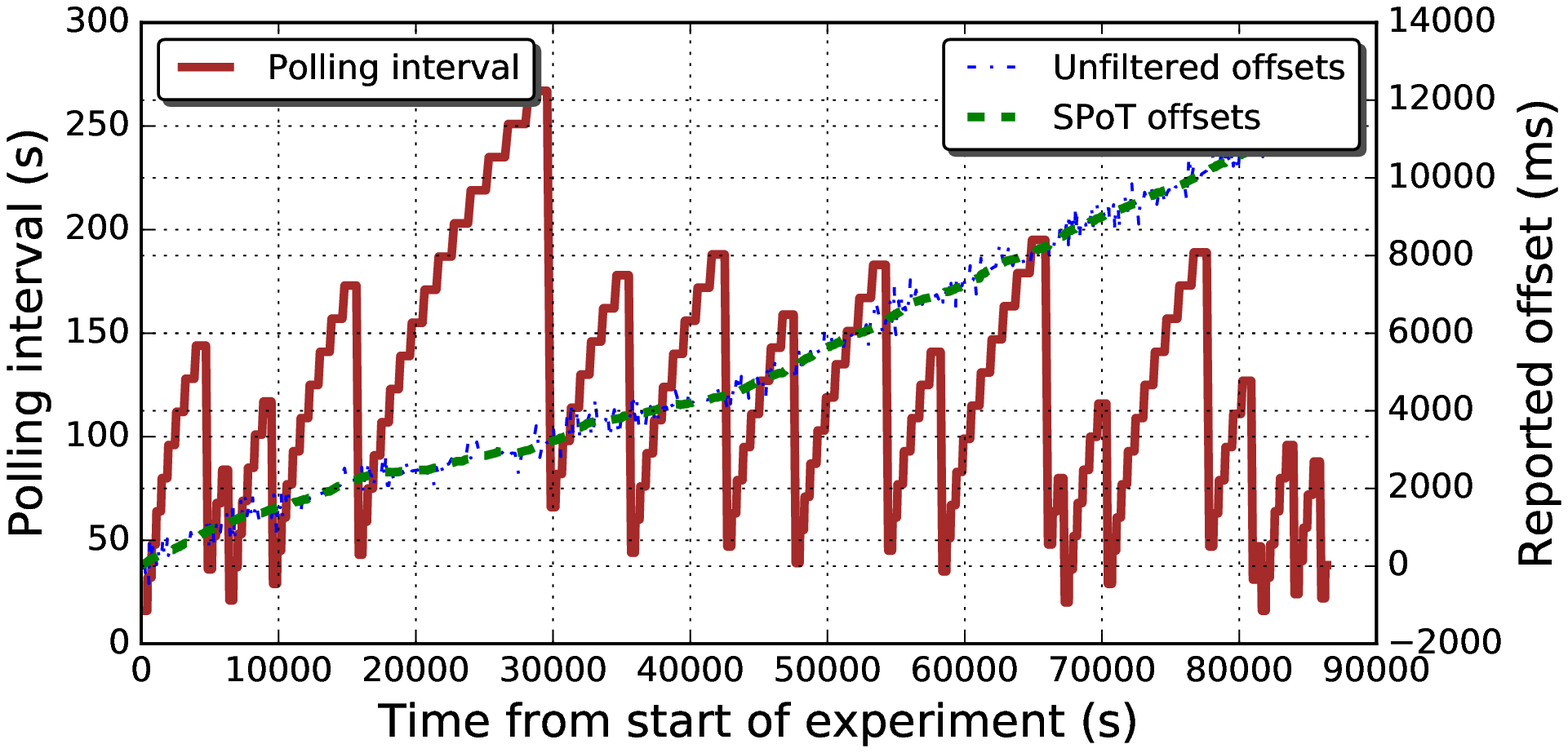}
  \includegraphics[width=14cm]{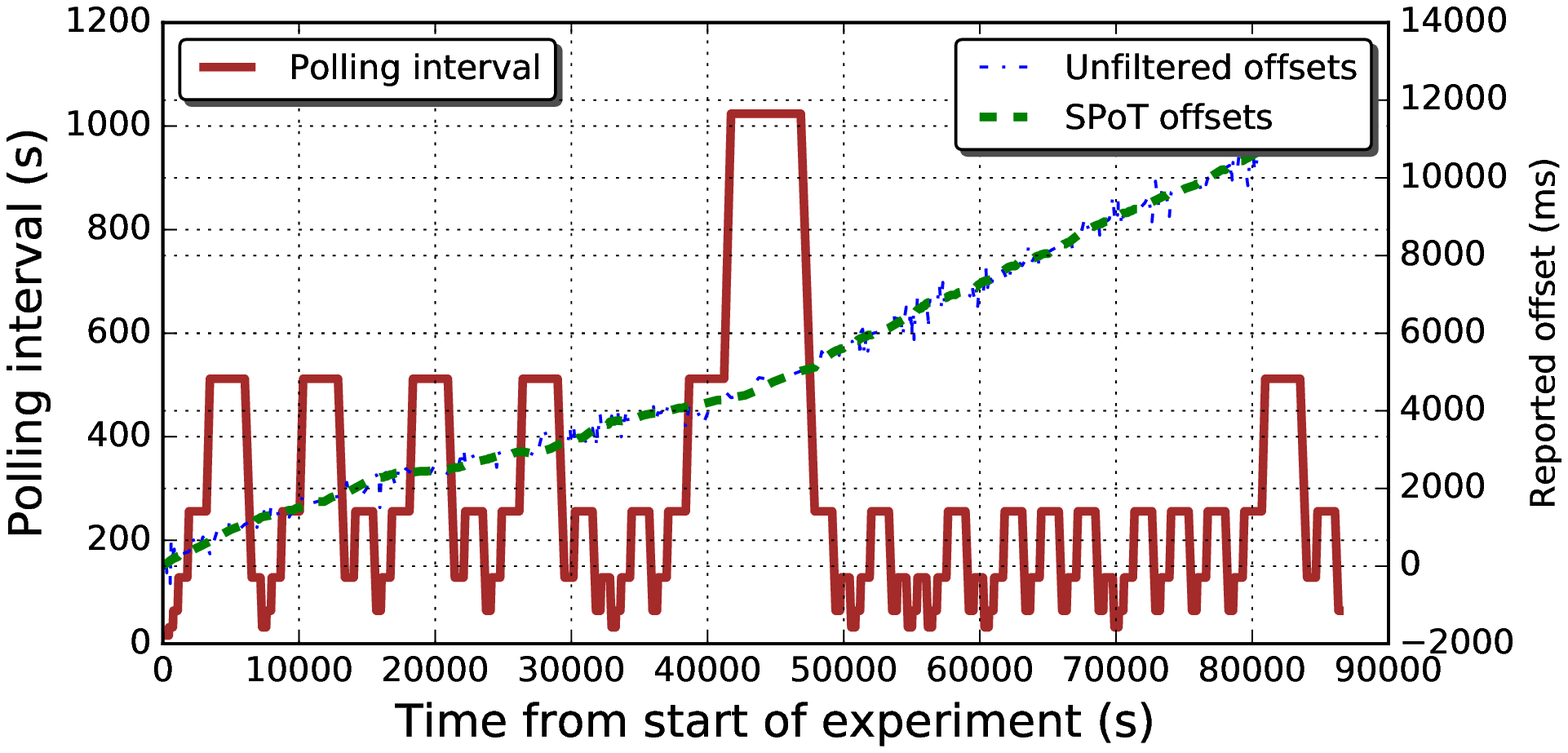}

  \caption{\label{fig:spotPollingBehavior}{\bf Comparison of polling
behaviors of SPoT on Arduino hardware with AIMD (top) and MIMD (bottom).}}
\end{figure*}
\vspace{0.4cm}

{\bf SPoT's polling behavior.} Figure~\ref{fig:spotPollingBehavior} compares the AIMD and MIMD polling behaviors of SPoT on the Arduino hardware. The figure shows that MIMD is more aggressive in increasing the polling interval. The RMSE error incurred by AIMD is 8.9 ms while the RMSE for MIMD is 14.7 ms. The number of offset measurements made by AIMD for a period of 24 hr is 953 and 545 for MIMD. This shows that IoT devices that have a strict energy budget but lower synchronization accuracy requirements should opt to use MIMD instead of AIMD.

\begin{figure*}[htb!]
  \centering
  \includegraphics[width=14cm]{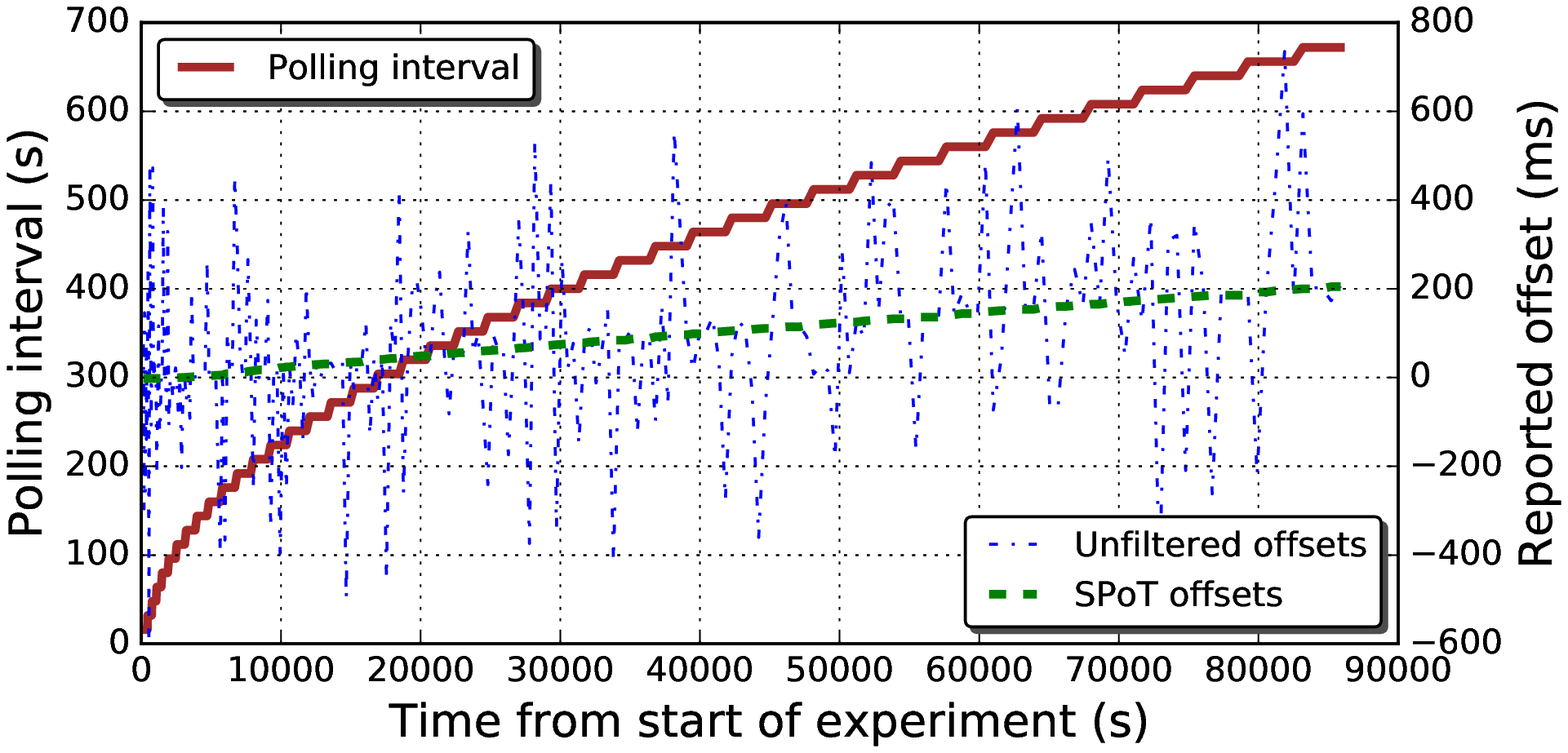}
  \includegraphics[width=14cm]{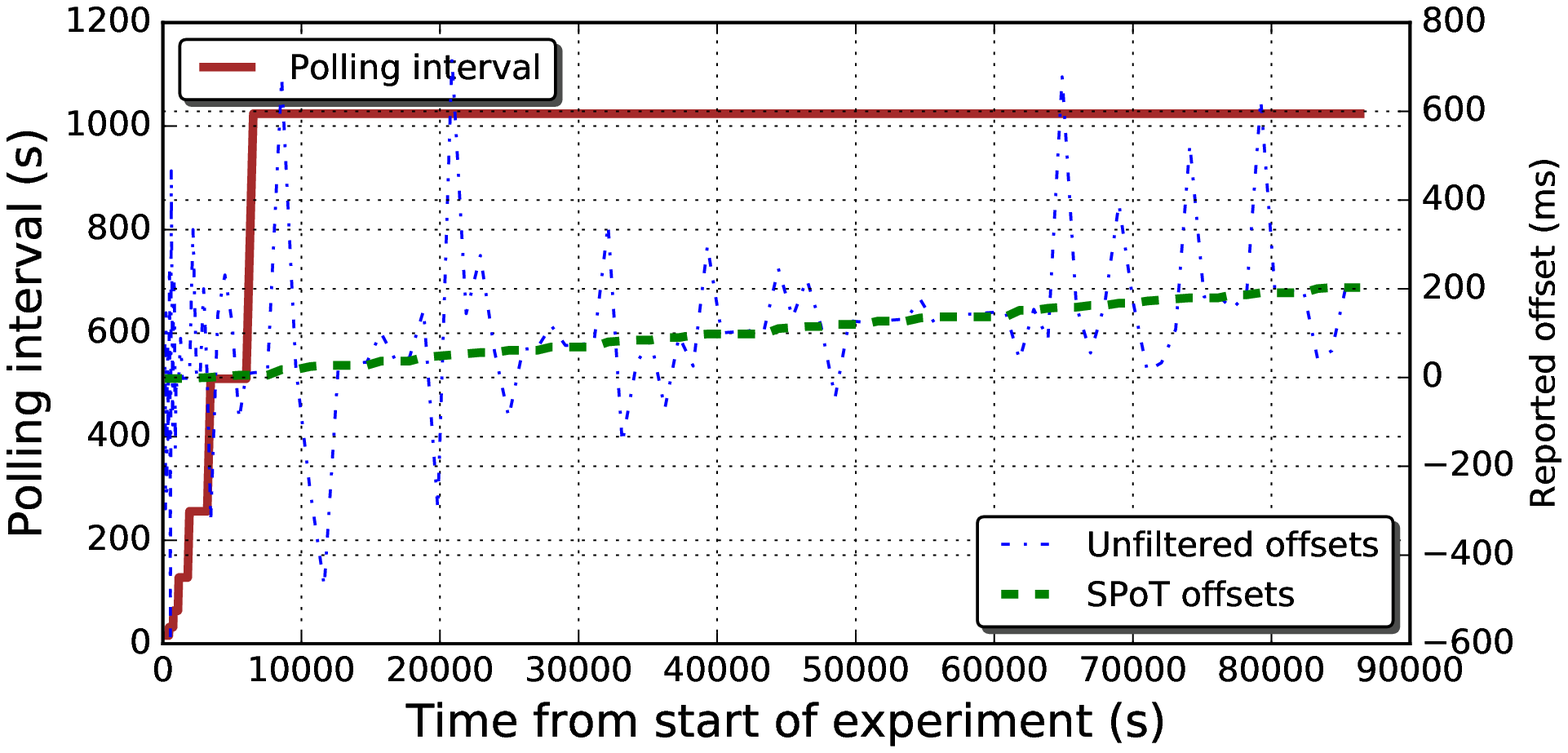}

  \caption{\label{fig:spotPollingBehaviorPi3}{\bf Comparison of polling behaviors of SPoT on Pi3 hardware with AIMD (top) and MIMD (bottom).}}
\end{figure*}
\vspace{0.4cm}

Similarly Figure~\ref{fig:spotPollingBehaviorPi3} shows the difference in behavior between AIMD and MIMD for SPoT running on the Raspberry Pi3.  The relatively stable clock hardware of the Pi3
can be seen from the total clock drift of about ~200 ms for a period of 24 hr compared to a drift of about ~12,000 ms on the Arduino hardware. As designed, SPoT is able to exploit the relatively higher stability of Pi3 hardware and reduce both offset and rate sync RMSE even in the presence of high noise. The RMSE for offset synchronization is 1.5 ms for AIMD and 3.0 ms for MIMD. Similarly, the rate synchronization RMSE values are 2.9 ms and 3.0 ms for AIMD and MIMD respectively. It is also clear that given the relative stability of Pi3, SPoT is effective in increasing the polling interval and particularly MIMD is able to reach the maximum polling value of 1024 s very rapidly so the number of offset measurements made by AIMD for a period of 24 hr is 269 and that of MIMD is only 132.

{\bf Scalability of SPoT.} We conduct a series of tests in the wide area to examine the scalability of SPoT. We note that our implementation uses a single node to serve all thin clients, which we use as a baseline. We use two cloud nodes, one on the east coast and another on the west coast of the U.S. One of them runs the SPoT server, while the other runs the client emulator. Since the SPoT server
maintains no state for thick clients its operation is essentially the same as the NTP reference server, thus we examine server scalability for thin clients only.
Both the server and the client emulator have their clock disciplined by NTP. 
That is, their expected offset throughout the experiment is 0 ms. We use several
benchmarking runs, each with 1, 10, 100, 1k, 5k, 10k and 15k clients. In each run, we
synchronize thin clients with the SPoT server for 5 minutes and calculate the average
RMSE of offsets.

\begin{table}[htb!]
\centering
\caption{{\bf RMSE errors for different number of clients in our scalability experiment.}}
\label{tab:scalability}
\begin{tabular}{c|c|c|c|c|c|c|c}
No. of thin clients & 1 & 10 & 100  & 1k & 5k & 10k & 15k\\ \hline
Avg. RMSE (ms) & 1.0  & 1.0  & 1.55 & 1.51 & 1.37 & 1.73 & 0.13\\
\end{tabular}
\end{table}

Table~\ref{tab:scalability} shows that synchronization accuracy remains consistent as the number of clients grows (avg. RMSE $<$ 2 ms). We note that the maximum number of thin clients considered in our experiments is an artifact of our prototype server implementation; we expect similarly consistent accuracy for larger numbers of clients if we deployed across multiple servers.



\begin{table}[htb!]
\centering
\caption{{\bf Memory and CPU profile of SpoT server.}}
\label{tab:resourceUsage}
\begin{tabular}{c|c|c|c|c|c|c|c}
No. of clients & 1 & 10 & 100  & 1k  & 10k & 100k & 1M\\ \hline
Instruction count & 669 k &  699 k & 719 k &	1 M & 4 M & 	34 M & 	332 M\\
Execution time (ms) & 0.02 &  0.02 & 0.04 &	0.18 & 0.72 & 	 6.53 & 	64.07\\
Memory (B) & 7 KB & 7 KB & 7 KB & 33 KB & 312 KB & 3 MB & 30 MB\\
\end{tabular}
\end{table}

To understand the resource consumption of SPoT we conduct micro-benchmark experiments that measure the CPU and the memory usage. 
Table~\ref{tab:resourceUsage} shows the average number of machine instructions and execution time (ms) required by the server to complete one round of synchronization for different numbers of clients 
when running on a Ubuntu 17.10 server with a quad-core 1.8 GHz Intel i5-3337U processor and 3.7 GiB memory, averaged over 100 runs. Total memory required by the SPoT server for different number of thin clients is also shown, which includes both the state information used by SPoT's synchronization algorithms as well as the book-keeping information required by the server to keep track of all clients. From the table we observe that the SPoT server's footprint is light on 
the CPU (execution time and instruction count) and memory usage, even for a high number of clients.

\begin{table}[htb!]
\centering
\caption{{\bf SPoT server throughput (PPS) required to support thin client.}}
\label{tab:throughput}
\begin{tabular}{c|c|c|c|c|c|c|c}
No. of clients & 1 & 10 & 100  & 1k  & 10k & 100k & 1M\\ \hline
Arduino-AIMD & 0.01 &	0.1 & 1 & 11 & 110 & 1.1 k & 11 k\\
Pi3-AIMD & 0.003 & 0.03 & 0.3 & 3 & 31 & 311 & 3.1 k\\
Arduino-MIMD & 0.006 & 0.06 & 0.63 &	6 & 63 & 630 &	6.3 k\\
Pi3-MIMD & 0.001 &	0.01 & 0.1 & 	1.5 & 15 & 152 & 1.5 k\\
\end{tabular}
\end{table}

Finally, using the number of packets exchanged by different polling methods of SPoT for Arduino devices and Pi3 for a period of 24 hr, we estimate the server throughput (packets/sec (PPS)) required to synchronize different number of thin clients. From Table~\ref{tab:throughput} we can see that the network overhead to run the SPoT server is low, with only a required throughput of about 6K PPS for 1M Arduino devices and about 1.5K PPS for stable hardware such as Pi3.

\section{Related work}
\label{sec:related_work_js} Our work relates most closely to prior studies that have examined environmental
effects such as temperature on oscillator performance and clock drift, as well
as synchronization protocols for sensor network platforms and other constrained
environments.  Regarding the first category of works, Schmid {\em et
al.}~\cite{schmid2008exploiting} extensively evaluate the problem of clock
drift in low-end oscillators in a variety of conditions.  They develop
a technique to address and compensate for drift by utilizing two oscillators
and exploiting subtle manufacturing differences between them.  In a somewhat
similar vein, there are a number of unpublished investigations by IoT
prototypers and hobbyists ({\em e.g.},~\cite{arduinoclock}) who provide
anecdotal evidence of the effects of environmental conditions on different
types of oscillators available on Arduino-based IoT platforms.  While the
effects of temperature on clock rate are well-known, these studies inform our
work by highlighting the challenges of addressing drift on low-end devices.


Wireless sensor networks (WSNs) typically consist of a large set of relatively homogenous nodes, with significant or extreme energy, bandwidth, and computational constraints.  While IoT devices are typically constrained in somewhat different ways than WSNs, there have been a number of time synchronization methods developed in the WSN context (Sundraraman {\em et al.} provide a survey of these methods in~\cite{sundararaman2005clock}) that have a bearing on our work, including~\cite{elson2002rbs,ganeriwal2003tpsn,marioti2004flooding,cox2005time,vanGreunen2003}. \begin{edittext}The protocols developed for WSN domain exploit characteristics of the broadcast MAC layer to avoid network inconsistencies that cause time synchronization errors. Hence, to synchronize to a global timescale such as UTC, these techniques require a UTC time source to be part of the same broadcast domain.\end{edittext}
More explicitly in the IoT domain, Sridhar {\em et
al.} describe the CheepSync time synchronization
protocol~\cite{sridhar2016cheepsync} which is designed to operate within
Bluetooth LE, exploiting the broadcast MAC in somewhat similar ways as
techniques in the WSN domain. Finally,  Kalman filters have been used in the context of time synchronization in order to model clock offset and skew, and to handle missing information~\cite{kim2012tracking,hamilton2008aces}.  These prior studies primarily used simulation as an evaluation technique, with some limited measurements from real oscillators. \begin{edittext}These methods that model clock hardware become highly challenging and resource intensive at scales introduced by the IoT domain, due to huge variability in drift characteristics exhibited by IoT hardware under different ambient temperature conditions, as discussed in \S \ref{temperature_experiments} and \S \ref{stability_experiments}.\end{edittext}

\section{Summary}
\label{sec:summary} In this paper, we consider the question of how to synchronize clocks in an
Internet of Things setting.  While clock synchronization has been considered
extensively in prior work, low-cost hardware and diverse environmental
deployments make IoT clock synchronization challenging.  We begin by
investigating clock drift in two standard prototyping platforms over a range of
operating conditions that would be typical for an IoT device.  We find clock
drift on the order of seconds over relatively short time periods.  This level
of variation makes standard protocols such as SNTP and those based on MQTT
ineffective.  We address this problem by developing a new system for
synchronizing clocks on IoT devices.  The system includes a lightweight client,
which is suitable for IoT devices with low processing power; a scalable
reference server that calculates clock offset and rate synchronization; and an
efficient packet exchange protocol called SPoT, which is also suitable for low
power devices.  We develop a prototype implementation of our system to evaluate
efficacy over a range of configurations, operating conditions and target
platforms. 
Our results show that SPoT outperforms MQTT and SNTP by a factor of 22 and 17
respectively, in the presence of high noise levels, and maintains a clock
accuracy of within 15ms at various noise levels. Finally, we report on the
scalability of our server implementation through microbenchmark experiments and
show that our system can scale to support large numbers of clients with minimal
resource utilization. In on-going work we plan to expand the range of devices
on which SPoT can be deployed, we plan to conduct tests in more diverse
configurations such as what might be found in home or shop floor deployments.

{
\balance

\bibliographystyle{ACM-Reference-Format}
\bibliography{paper}


\begin{thebibliography}{59}


\ifx \showCODEN    \undefined \def \showCODEN     #1{\unskip}     \fi
\ifx \showDOI      \undefined \def \showDOI       #1{#1}\fi
\ifx \showISBNx    \undefined \def \showISBNx     #1{\unskip}     \fi
\ifx \showISBNxiii \undefined \def \showISBNxiii  #1{\unskip}     \fi
\ifx \showISSN     \undefined \def \showISSN      #1{\unskip}     \fi
\ifx \showLCCN     \undefined \def \showLCCN      #1{\unskip}     \fi
\ifx \shownote     \undefined \def \shownote      #1{#1}          \fi
\ifx \showarticletitle \undefined \def \showarticletitle #1{#1}   \fi
\ifx \showURL      \undefined \def \showURL       {\relax}        \fi
\providecommand\bibfield[2]{#2}
\providecommand\bibinfo[2]{#2}
\providecommand\natexlab[1]{#1}
\providecommand\showeprint[2][]{arXiv:#2}

\bibitem[\protect\citeauthoryear{??}{aws}{[n. d.]}]%
        {awsiot}
 \bibinfo{year}{[n. d.]}\natexlab{}.
\newblock \bibinfo{title}{{Amazon AWS IoT}}.
\newblock \bibinfo{howpublished}{\url{https://aws.amazon.com/iot/}}.
\newblock


\bibitem[\protect\citeauthoryear{??}{asi}{[n. d.]}]%
        {asioLib}
 \bibinfo{year}{[n. d.]}\natexlab{}.
\newblock \bibinfo{title}{{Asio C++ library.}}
\newblock \bibinfo{howpublished}{\url{https://think-async.com/}}.
\newblock


\bibitem[\protect\citeauthoryear{??}{par}{[n. d.]}]%
        {paritySNTP}
 \bibinfo{year}{[n. d.]}\natexlab{}.
\newblock \bibinfo{title}{{Fast, light, robust Ethereum implementation.}}
\newblock
  \bibinfo{howpublished}{\url{https://github.com/paritytech/parity/blob/master/dapps/node-health/src/time.rs\#L22-L31}}.
\newblock


\bibitem[\protect\citeauthoryear{??}{gri}{[n. d.]}]%
        {grillo}
 \bibinfo{year}{[n. d.]}\natexlab{}.
\newblock \bibinfo{title}{{IoT: Sensing Earthquakes before hand with Grillo}}.
\newblock
  \bibinfo{howpublished}{\url{https://electronicsofthings.com/expert-opinion/iot-sensing-earthquakes-before-hand-with-grillo/}}.
\newblock


\bibitem[\protect\citeauthoryear{??}{ard}{[n. d.]}]%
        {arduinoFamous}
 \bibinfo{year}{[n. d.]}\natexlab{}.
\newblock \bibinfo{title}{{Maker Madness: The Best IoT Boards of 2016.}}
\newblock
  \bibinfo{howpublished}{\url{https://blog.hackster.io/maker-madness-the-best-iot-boards-of-2016-cfc2382daf64}}.
\newblock


\bibitem[\protect\citeauthoryear{??}{azu}{[n. d.]}]%
        {azureiot}
 \bibinfo{year}{[n. d.]}\natexlab{}.
\newblock \bibinfo{title}{{Microsoft Azure IoT}}.
\newblock
  \bibinfo{howpublished}{\url{https://azure.microsoft.com/en-us/suites/iot-suite/}}.
\newblock


\bibitem[\protect\citeauthoryear{??}{pus}{[n. d.]}]%
        {pushEpoch}
 \bibinfo{year}{[n. d.]}\natexlab{}.
\newblock \bibinfo{title}{{MQTT Time Utilities.}}
\newblock
  \bibinfo{howpublished}{\url{https://io.adafruit.com/blog/feature/2016/06/01/time-utilities/}}.
\newblock


\bibitem[\protect\citeauthoryear{??}{ntp}{[n. d.]a}]%
        {ntpClockDiscipline}
 \bibinfo{year}{[n. d.]}\natexlab{a}.
\newblock \bibinfo{title}{{NTP Clock Discipline Algorithm.}}
\newblock
  \bibinfo{howpublished}{\url{https://www.eecis.udel.edu/~mills/ntp/html/discipline.html}}.
\newblock


\bibitem[\protect\citeauthoryear{??}{ntp}{[n. d.]b}]%
        {ntpFilter}
 \bibinfo{year}{[n. d.]}\natexlab{b}.
\newblock \bibinfo{title}{{NTP Clock Filter Algorithm.}}
\newblock
  \bibinfo{howpublished}{\url{https://www.eecis.udel.edu/~mills/ntp/html/filter.html}}.
\newblock


\bibitem[\protect\citeauthoryear{??}{ntp}{[n. d.]c}]%
        {ntpHPFilter}
 \bibinfo{year}{[n. d.]}\natexlab{c}.
\newblock \bibinfo{title}{{NTP the Huff-n'-Puff filter.}}
\newblock
  \bibinfo{howpublished}{\url{https://www.eecis.udel.edu/~mills/ntp/html/huffpuff.html}}.
\newblock


\bibitem[\protect\citeauthoryear{??}{pul}{[n. d.]}]%
        {pullRest}
 \bibinfo{year}{[n. d.]}\natexlab{}.
\newblock \bibinfo{title}{{Particle Docs.}}
\newblock
  \bibinfo{howpublished}{\url{https://docs.particle.io/reference/firmware/electron}}.
\newblock


\bibitem[\protect\citeauthoryear{??}{rfc}{1983}]%
        {rfc868}
 \bibinfo{year}{1983}\natexlab{}.
\newblock \bibinfo{title}{{Time Protocol}}.
\newblock \bibinfo{howpublished}{\url{https://tools.ietf.org/html/rfc868}}.
\newblock


\bibitem[\protect\citeauthoryear{??}{rfc}{1999}]%
        {rfc2616}
 \bibinfo{year}{1999}\natexlab{}.
\newblock \bibinfo{title}{{Hypertext Transfer Protocol (HTTP)}}.
\newblock \bibinfo{howpublished}{\url{https://tools.ietf.org/html/rfc2616}}.
\newblock


\bibitem[\protect\citeauthoryear{??}{ios}{2010}]%
        {iosSNTP}
 \bibinfo{year}{2010}\natexlab{}.
\newblock \bibinfo{title}{{SNTP implementation for iOS}}.
\newblock \bibinfo{howpublished}{\url{https://github.com/jbenet/ios-ntp}}.
\newblock


\bibitem[\protect\citeauthoryear{??}{rfc}{2011}]%
        {rfc6455}
 \bibinfo{year}{2011}\natexlab{}.
\newblock \bibinfo{title}{{The WebSocket Protocol (WebSockets)}}.
\newblock \bibinfo{howpublished}{\url{https://tools.ietf.org/html/rfc6455}}.
\newblock


\bibitem[\protect\citeauthoryear{??}{amq}{2012}]%
        {amqp}
 \bibinfo{year}{2012}\natexlab{}.
\newblock \bibinfo{title}{{Advanced Message Queuing Protocol (AMQP)}}.
\newblock
  \bibinfo{howpublished}{\url{http://docs.oasis-open.org/amqp/core/v1.0/amqp-core-complete-v1.0.pdf}}.
\newblock


\bibitem[\protect\citeauthoryear{??}{mqt}{2014}]%
        {mqtt}
 \bibinfo{year}{2014}\natexlab{}.
\newblock \bibinfo{title}{{MQ Telemetry Transport (MQTT)}}.
\newblock
  \bibinfo{howpublished}{\url{http://docs.oasis-open.org/mqtt/mqtt/v3.1.1/os/mqtt-v3.1.1-os.html}}.
\newblock


\bibitem[\protect\citeauthoryear{??}{ntp}{2014}]%
        {ntpd}
 \bibinfo{year}{2014}\natexlab{}.
\newblock \bibinfo{title}{{NTP Daemon.}}
\newblock
  \bibinfo{howpublished}{\url{https://www.eecis.udel.edu/~mills/ntp/html/ntpd.html}}.
\newblock


\bibitem[\protect\citeauthoryear{??}{sna}{2016}]%
        {snapchatSNTP}
 \bibinfo{year}{2016}\natexlab{}.
\newblock \bibinfo{title}{{Snapchat coding error nearly destroys all of time
  for the internet}}.
\newblock
  \bibinfo{howpublished}{\url{https://www.theregister.co.uk/2016/12/21/snapchat_coding_error_nearly_destroys_all_of_time_for_the_internet/}}.
\newblock


\bibitem[\protect\citeauthoryear{??}{ard}{2017}]%
        {arduinomkr1000}
 \bibinfo{year}{2017}\natexlab{}.
\newblock \bibinfo{title}{{Arduino MKR1000}}.
\newblock
  \bibinfo{howpublished}{\url{https://www.arduino.cc/en/Main/ArduinoMKR1000}}.
\newblock


\bibitem[\protect\citeauthoryear{??}{eth}{2017}]%
        {ethereumSyncError}
 \bibinfo{year}{2017}\natexlab{}.
\newblock \bibinfo{title}{{How do I set the time to be synchronized on
  Parity?}}
\newblock
  \bibinfo{howpublished}{\url{https://ethereum.stackexchange.com/questions/23599/how-do-i-set-the-time-to-be-synchronized-on-parity}}.
\newblock


\bibitem[\protect\citeauthoryear{??}{ntp}{2017}]%
        {ntpdate}
 \bibinfo{year}{2017}\natexlab{}.
\newblock \bibinfo{title}{{ntpdate Documentation}}.
\newblock \bibinfo{howpublished}{\url{http://doc.ntp.org/4.1.1/ntpdate.htm}}.
\newblock


\bibitem[\protect\citeauthoryear{??}{par}{2017a}]%
        {particleThreading}
 \bibinfo{year}{2017}\natexlab{a}.
\newblock \bibinfo{title}{{Particle Docs - system threads.}}
\newblock
  \bibinfo{howpublished}{\url{https://docs.particle.io/reference/firmware/photon/\#system-thread}}.
\newblock


\bibitem[\protect\citeauthoryear{??}{ras}{2017}]%
        {raspberrypi3}
 \bibinfo{year}{2017}\natexlab{}.
\newblock \bibinfo{title}{{Raspberry Pi3}}.
\newblock
  \bibinfo{howpublished}{\url{https://www.raspberrypi.org/products/raspberry-pi-3-model-b/}}.
\newblock


\bibitem[\protect\citeauthoryear{??}{sha}{2017}]%
        {shakeAlert}
 \bibinfo{year}{2017}\natexlab{}.
\newblock \bibinfo{title}{{ShakeAlert: Implementing Public Earthquake Early
  Warning for the U.S.}}
\newblock
  \bibinfo{howpublished}{\url{https://www.wmo.int/pages/prog/drr/documents/mhews-ref/posters-pdfs/2.49
  - Given D ShakeAlert MHEWC 2017 poster.pdf}}.
\newblock


\bibitem[\protect\citeauthoryear{??}{par}{2017b}]%
        {paritySyncError}
 \bibinfo{year}{2017}\natexlab{b}.
\newblock \bibinfo{title}{{Your clock is not in sync}}.
\newblock
  \bibinfo{howpublished}{\url{https://github.com/paritytech/parity/issues/6684}}.
\newblock


\bibitem[\protect\citeauthoryear{??}{zig}{2017}]%
        {zigbee}
 \bibinfo{year}{2017}\natexlab{}.
\newblock \bibinfo{title}{{ZigBee Specification}}.
\newblock
  \bibinfo{howpublished}{\url{http://www.zigbee.org/zigbee-for-developers/network-specifications/}}.
\newblock


\bibitem[\protect\citeauthoryear{??}{iot}{2018a}]%
        {iotBlockchain}
 \bibinfo{year}{2018}\natexlab{a}.
\newblock \bibinfo{title}{{Blockchain IoT - IBM Watson IoT}}.
\newblock
  \bibinfo{howpublished}{\url{https://www.ibm.com/internet-of-things/spotlight/blockchain}}.
\newblock


\bibitem[\protect\citeauthoryear{??}{hyp}{2018}]%
        {hyperLedger}
 \bibinfo{year}{2018}\natexlab{}.
\newblock \bibinfo{title}{{Hyperledger - Open source blockchain for businesses
  - IBM Blockchain}}.
\newblock
  \bibinfo{howpublished}{\url{https://www.ibm.com/blockchain/hyperledger.html}}.
\newblock


\bibitem[\protect\citeauthoryear{??}{ble}{2018}]%
        {ble}
 \bibinfo{year}{2018}\natexlab{}.
\newblock \bibinfo{title}{{Radio Versions}}.
\newblock
  \bibinfo{howpublished}{\url{https://www.bluetooth.com/bluetooth-technology/radio-versions}}.
\newblock


\bibitem[\protect\citeauthoryear{??}{iot}{2018b}]%
        {iotBlockchainSecurity}
 \bibinfo{year}{2018}\natexlab{b}.
\newblock \bibinfo{title}{{Using blockchain to secure the internet of things}}.
\newblock
  \bibinfo{howpublished}{\url{https://theconversation.com/using-blockchain-to-secure-the-internet-of-things-90002}}.
\newblock


\bibitem[\protect\citeauthoryear{Cox, Jovanov, and Milenkovic}{Cox
  et~al\mbox{.}}{2005}]%
        {cox2005time}
\bibfield{author}{\bibinfo{person}{D. Cox}, \bibinfo{person}{E. Jovanov}, {and}
  \bibinfo{person}{A. Milenkovic}.} \bibinfo{year}{2005}\natexlab{}.
\newblock \showarticletitle{{Time Synchronization for ZigBee Networks}}. In
  \bibinfo{booktitle}{\emph{In IEEE SSST}}.
\newblock


\bibitem[\protect\citeauthoryear{Elson et~al\mbox{.}}{Elson
  et~al\mbox{.}}{2002}]%
        {elson2002rbs}
\bibfield{author}{\bibinfo{person}{J. Elson} {et~al\mbox{.}}}
  \bibinfo{year}{2002}\natexlab{}.
\newblock \showarticletitle{{Fine-grained Network Time Synchronization Using
  Reference Broadcasts}}. In \bibinfo{booktitle}{\emph{Usenix OSDI}}.
\newblock


\bibitem[\protect\citeauthoryear{Ganeriwal et~al\mbox{.}}{Ganeriwal
  et~al\mbox{.}}{2003}]%
        {ganeriwal2003tpsn}
\bibfield{author}{\bibinfo{person}{S. Ganeriwal} {et~al\mbox{.}}}
  \bibinfo{year}{2003}\natexlab{}.
\newblock \showarticletitle{{Timing-sync Protocol for Sensor Networks}}. In
  \bibinfo{booktitle}{\emph{ACM SenSys}}.
\newblock


\bibitem[\protect\citeauthoryear{Hamilton, Ma, Zhao, and Xu}{Hamilton
  et~al\mbox{.}}{2008}]%
        {hamilton2008aces}
\bibfield{author}{\bibinfo{person}{B.R. Hamilton}, \bibinfo{person}{X. Ma},
  \bibinfo{person}{Q. Zhao}, {and} \bibinfo{person}{J. Xu}.}
  \bibinfo{year}{2008}\natexlab{}.
\newblock \showarticletitle{{ACES: Adaptive Clock Estimation and
  Synchronization using Kalman Filtering}}. In \bibinfo{booktitle}{\emph{ACM
  Mobicom}}.
\newblock


\bibitem[\protect\citeauthoryear{{IEEE}}{{IEEE}}{2008}]%
        {ieeeptp}
\bibfield{author}{\bibinfo{person}{{IEEE}}.} \bibinfo{year}{2008}\natexlab{}.
\newblock \bibinfo{title}{{IEEE 1588 Precision Time Protocol (PTP), Version 2
  Specification}}.
\newblock
\newblock


\bibitem[\protect\citeauthoryear{Kim, Ma, and Hamilton}{Kim
  et~al\mbox{.}}{2012}]%
        {kim2012tracking}
\bibfield{author}{\bibinfo{person}{H. Kim}, \bibinfo{person}{X. Ma}, {and}
  \bibinfo{person}{B.R. Hamilton}.} \bibinfo{year}{2012}\natexlab{}.
\newblock \showarticletitle{{Tracking Low-precision Clocks with Time-varying
  Drifts using Kalman Filtering}}.
\newblock \bibinfo{journal}{\emph{IEEE/ACM TON}} (\bibinfo{year}{2012}).
\newblock


\bibitem[\protect\citeauthoryear{Lee, Wang, Shrivastav, and Weatherspoon}{Lee
  et~al\mbox{.}}{2016}]%
        {lee2016globally}
\bibfield{author}{\bibinfo{person}{K.S. Lee}, \bibinfo{person}{H. Wang},
  \bibinfo{person}{V. Shrivastav}, {and} \bibinfo{person}{H. Weatherspoon}.}
  \bibinfo{year}{2016}\natexlab{}.
\newblock \showarticletitle{{Globally Synchronized Time via Datacenter
  Networks}}. In \bibinfo{booktitle}{\emph{ACM SIGCOMM}}.
\newblock


\bibitem[\protect\citeauthoryear{Levine}{Levine}{2016}]%
        {syncUnstableDelay16}
\bibfield{author}{\bibinfo{person}{J. Levine}.}
  \bibinfo{year}{2016}\natexlab{}.
\newblock \bibinfo{journal}{\emph{IEEE Trans Ultrason Ferroelectr Freq
  Control}}  \bibinfo{volume}{63} (\bibinfo{date}{Jan-04-2016}
  \bibinfo{year}{2016}), \bibinfo{pages}{561 -- 570}.
\newblock
\showISSN{0885-3010}
\urldef\tempurl%
\url{https://doi.org/10.1109/TUFFC.2015.2495014}
\showDOI{\tempurl}


\bibitem[\protect\citeauthoryear{Mani, Durairajan, Barford, and Sommers}{Mani
  et~al\mbox{.}}{2016}]%
        {Mani2016}
\bibfield{author}{\bibinfo{person}{S.K. Mani}, \bibinfo{person}{R. Durairajan},
  \bibinfo{person}{P. Barford}, {and} \bibinfo{person}{J. Sommers}.}
  \bibinfo{year}{2016}\natexlab{}.
\newblock \showarticletitle{{MNTP: Enhancing Time Synchronization for Mobile
  Devices}}. In \bibinfo{booktitle}{\emph{ACM IMC}}.
\newblock


\bibitem[\protect\citeauthoryear{Mar{\'o}ti, Kusy, Simon, and
  L{\'e}deczi}{Mar{\'o}ti et~al\mbox{.}}{2004}]%
        {marioti2004flooding}
\bibfield{author}{\bibinfo{person}{M Mar{\'o}ti}, \bibinfo{person}{B Kusy},
  \bibinfo{person}{G Simon}, {and} \bibinfo{person}{{\'A} L{\'e}deczi}.}
  \bibinfo{year}{2004}\natexlab{}.
\newblock \showarticletitle{{The Flooding Time Synchronization Protocol}}. In
  \bibinfo{booktitle}{\emph{ACM SenSys}}.
\newblock


\bibitem[\protect\citeauthoryear{Marzullo and Owicki}{Marzullo and
  Owicki}{1983}]%
        {Marzullo83}
\bibfield{author}{\bibinfo{person}{K. Marzullo} {and} \bibinfo{person}{S.
  Owicki}.} \bibinfo{year}{1983}\natexlab{}.
\newblock \showarticletitle{{Maintaining the Time in a Distributed System}}. In
  \bibinfo{booktitle}{\emph{ACM PODC}}.
\newblock


\bibitem[\protect\citeauthoryear{Membrey, Veitch, and Chang}{Membrey
  et~al\mbox{.}}{2016}]%
        {timeToMeasurePi}
\bibfield{author}{\bibinfo{person}{P. Membrey}, \bibinfo{person}{D. Veitch},
  {and} \bibinfo{person}{R.K.C. Chang}.} \bibinfo{year}{2016}\natexlab{}.
\newblock \showarticletitle{{Time to Measure the Pi}}. In
  \bibinfo{booktitle}{\emph{ACM IMC}}.
\newblock
\urldef\tempurl%
\url{https://doi.org/10.1145/2987443.2987476}
\showDOI{\tempurl}


\bibitem[\protect\citeauthoryear{Mills}{Mills}{1981}]%
        {mills81dcnet}
\bibfield{author}{\bibinfo{person}{D.L. Mills}.}
  \bibinfo{year}{1981}\natexlab{}.
\newblock \bibinfo{title}{{DCNET Internet Clock Service}}.
\newblock \bibinfo{howpublished}{\url{https://tools.ietf.org/html/rfc778}}.
\newblock


\bibitem[\protect\citeauthoryear{Mills}{Mills}{1985a}]%
        {mills85alg}
\bibfield{author}{\bibinfo{person}{D.L. Mills}.}
  \bibinfo{year}{1985}\natexlab{a}.
\newblock \bibinfo{title}{{Algorithms for Synchronizing Network Clocks}}.
\newblock \bibinfo{howpublished}{\url{https://tools.ietf.org/html/rfc956}}.
\newblock


\bibitem[\protect\citeauthoryear{Mills}{Mills}{1985b}]%
        {rfc958}
\bibfield{author}{\bibinfo{person}{D.L. Mills}.}
  \bibinfo{year}{1985}\natexlab{b}.
\newblock \bibinfo{title}{{Network Time Protocol (NTP)}}.
\newblock \bibinfo{howpublished}{\url{https://tools.ietf.org/html/rfc958}}.
\newblock


\bibitem[\protect\citeauthoryear{Mills}{Mills}{1995}]%
        {rfc1769}
\bibfield{author}{\bibinfo{person}{D.L. Mills}.}
  \bibinfo{year}{1995}\natexlab{}.
\newblock \bibinfo{title}{{Simple Network Time Protocol (SNTP)}}.
\newblock \bibinfo{howpublished}{\url{https://tools.ietf.org/html/rfc1769}}.
\newblock


\bibitem[\protect\citeauthoryear{Mills}{Mills}{1996}]%
        {mills1996networkAllan}
\bibfield{author}{\bibinfo{person}{D.L. Mills}.}
  \bibinfo{year}{1996}\natexlab{}.
\newblock \bibinfo{booktitle}{\emph{The network computer as precision
  timekeeper}}.
\newblock \bibinfo{type}{{T}echnical {R}eport}. \bibinfo{institution}{DELAWARE
  UNIV NEWARK DEPT OF ELECTRICAL ENGINEERING}.
\newblock


\bibitem[\protect\citeauthoryear{Mills}{Mills}{1998}]%
        {ntpClockDiscipline98}
\bibfield{author}{\bibinfo{person}{D.L. Mills}.}
  \bibinfo{year}{1998}\natexlab{}.
\newblock \showarticletitle{Adaptive hybrid clock discipline algorithm for the
  network time protocol}.
\newblock \bibinfo{journal}{\emph{IEEE/ACM Trans. Netw.}} \bibinfo{volume}{6},
  \bibinfo{number}{5} (\bibinfo{date}{Oct} \bibinfo{year}{1998}),
  \bibinfo{pages}{505--514}.
\newblock


\bibitem[\protect\citeauthoryear{Moon, Skelly, and Towsley}{Moon
  et~al\mbox{.}}{1999}]%
        {moon1999estimation}
\bibfield{author}{\bibinfo{person}{S.B. Moon}, \bibinfo{person}{P. Skelly},
  {and} \bibinfo{person}{D. Towsley}.} \bibinfo{year}{1999}\natexlab{}.
\newblock \showarticletitle{Estimation and removal of clock skew from network
  delay measurements}. In \bibinfo{booktitle}{\emph{Proceedings of
  INFOCOM'99}}.
\newblock


\bibitem[\protect\citeauthoryear{Qu et~al\mbox{.}}{Qu et~al\mbox{.}}{2016}]%
        {Qu2016}
\bibfield{author}{\bibinfo{person}{Lei S.P. Wang~Z.Z. Qu, T.} {et~al\mbox{.}}}
  \bibinfo{year}{2016}\natexlab{}.
\newblock \bibinfo{journal}{\emph{Int J Adv Manuf Technol}}
  \bibinfo{volume}{84}, \bibinfo{number}{1} (\bibinfo{date}{01 Apr}
  \bibinfo{year}{2016}), \bibinfo{pages}{147--164}.
\newblock
\showISSN{1433-3015}
\urldef\tempurl%
\url{https://doi.org/10.1007/s00170-015-7220-1}
\showDOI{\tempurl}


\bibitem[\protect\citeauthoryear{Ridoux, Veitch, and Broomhead}{Ridoux
  et~al\mbox{.}}{2012}]%
        {ridoux2012case}
\bibfield{author}{\bibinfo{person}{J. Ridoux}, \bibinfo{person}{D. Veitch},
  {and} \bibinfo{person}{T. Broomhead}.} \bibinfo{year}{2012}\natexlab{}.
\newblock \showarticletitle{The case for feed-forward clock synchronization}.
\newblock \bibinfo{journal}{\emph{IEEE/ACM Transactions on Networking (TON)}}
  \bibinfo{volume}{20}, \bibinfo{number}{1} (\bibinfo{year}{2012}),
  \bibinfo{pages}{231--242}.
\newblock


\bibitem[\protect\citeauthoryear{Schmid et~al\mbox{.}}{Schmid
  et~al\mbox{.}}{2008}]%
        {schmid2008exploiting}
\bibfield{author}{\bibinfo{person}{T. Schmid} {et~al\mbox{.}}}
  \bibinfo{year}{2008}\natexlab{}.
\newblock \showarticletitle{{Exploiting Manufacturing Variations for
  Compensating Environment-induced Clock Drift in Time Synchronization}}.
\newblock \bibinfo{journal}{\emph{ACM SIGMETRICS}} (\bibinfo{year}{2008}).
\newblock


\bibitem[\protect\citeauthoryear{Sridhar, Misra, Gill, and Warrior}{Sridhar
  et~al\mbox{.}}{2016}]%
        {sridhar2016cheepsync}
\bibfield{author}{\bibinfo{person}{S. Sridhar}, \bibinfo{person}{P. Misra},
  \bibinfo{person}{G.S. Gill}, {and} \bibinfo{person}{J. Warrior}.}
  \bibinfo{year}{2016}\natexlab{}.
\newblock \showarticletitle{{CheepSync: A Time Synchronization Service for
  Resource Constrained Bluetooth LE Advertisers}}.
\newblock \bibinfo{journal}{\emph{IEEE Communications}} (\bibinfo{year}{2016}).
\newblock


\bibitem[\protect\citeauthoryear{Sundararaman et~al\mbox{.}}{Sundararaman
  et~al\mbox{.}}{2005}]%
        {sundararaman2005clock}
\bibfield{author}{\bibinfo{person}{B. Sundararaman} {et~al\mbox{.}}}
  \bibinfo{year}{2005}\natexlab{}.
\newblock \showarticletitle{Clock Synchronization for Wireless Sensor Networks:
  A Survey}.
\newblock \bibinfo{journal}{\emph{Adhoc networks}} (\bibinfo{year}{2005}).
\newblock


\bibitem[\protect\citeauthoryear{van Greunen and Rabaey}{van Greunen and
  Rabaey}{2003}]%
        {vanGreunen2003}
\bibfield{author}{\bibinfo{person}{J. van Greunen} {and} \bibinfo{person}{J.
  Rabaey}.} \bibinfo{year}{2003}\natexlab{}.
\newblock \showarticletitle{{Lightweight Time Synchronization for Sensor
  Networks}}. In \bibinfo{booktitle}{\emph{ACM WSNA}}.
\newblock


\bibitem[\protect\citeauthoryear{van Rantwijk}{van Rantwijk}{2017}]%
        {arduinoclock}
\bibfield{author}{\bibinfo{person}{J. van Rantwijk}.}
  \bibinfo{year}{2017}\natexlab{}.
\newblock \bibinfo{title}{Arduino clock frequency accuracy}.
\newblock
  \bibinfo{howpublished}{\url{http://jorisvr.nl/article/arduino-frequency}}.
\newblock


\bibitem[\protect\citeauthoryear{Veitch, Babu, and P\`{a}sztor}{Veitch
  et~al\mbox{.}}{2004}]%
        {radClock04}
\bibfield{author}{\bibinfo{person}{D. Veitch}, \bibinfo{person}{S. Babu}, {and}
  \bibinfo{person}{A. P\`{a}sztor}.} \bibinfo{year}{2004}\natexlab{}.
\newblock \showarticletitle{{Robust Synchronization of Software Clocks Across
  the Internet}}. In \bibinfo{booktitle}{\emph{ACM IMC}}.
\newblock
\urldef\tempurl%
\url{https://doi.org/10.1145/1028788.1028817}
\showDOI{\tempurl}


\bibitem[\protect\citeauthoryear{Veitch, Ridoux, and Korada}{Veitch
  et~al\mbox{.}}{2009}]%
        {veitch2009robust}
\bibfield{author}{\bibinfo{person}{D. Veitch}, \bibinfo{person}{J. Ridoux},
  {and} \bibinfo{person}{S.B. Korada}.} \bibinfo{year}{2009}\natexlab{}.
\newblock \showarticletitle{Robust synchronization of absolute and difference
  clocks over networks}.
\newblock \bibinfo{journal}{\emph{IEEE/ACM Transactions on Networking (TON)}}
  \bibinfo{volume}{17}, \bibinfo{number}{2} (\bibinfo{year}{2009}),
  \bibinfo{pages}{417--430}.
\newblock


\end{thebibliography}
}

\end{document}